\documentclass[10pt,journal,compsoc]{IEEEtran}
\ifCLASSOPTIONcompsoc
  \usepackage[nocompress]{cite}
\else
  \usepackage{cite}
\fi
\usepackage[hyphens]{url}

\usepackage{graphicx}
\usepackage{subfiles}
\usepackage{booktabs}
\usepackage{adjustbox}
\usepackage{algorithm}
\usepackage[]{algpseudocode}
\usepackage{amsmath,mathtools} 
\usepackage{hyperref}

\usepackage{tikz}
\newcommand*\circled[1]{\tikz[baseline=(char.base)]{
            \node[shape=circle,draw,inner sep=2pt] (char) {#1};}}

\def\BibTeX{{\rm B\kern-.05em{\sc i\kern-.025em b}\kern-.08em
    T\kern-.1667em\lower.7ex\hbox{E}\kern-.125emX}}

\begin{document}
%
\title{Dynamic Sampling Rate: Harnessing Frame Coherence in Graphics Applications for Energy-Efficient GPUs}
%
%
%
%

\author{Martí Anglada, Enrique de Lucas, Joan-Manuel Parcerisa, Juan L. Aragón,
        and~Antonio~González,~\IEEEmembership{Fellow,~IEEE}
\IEEEcompsocitemizethanks{\IEEEcompsocthanksitem M. Anglada, J.-M. Parcerisa and A. Gonz\'{a}lez are with the Departament d'Arquitectura de Computadors, Universitat Polit\`{e}cnica de Catalunya, c/Jordi Girona 1-3, Barcelona 08034, Spain. \protect\\
E-mail: \{manglada, jmanel, antonio\}@ac.upc.edu
\IEEEcompsocthanksitem Enrique de Lucas was with Esperanto Technologies, Mountain View, CA 94040, United States. He is now with Imagination Technologies, Imagination House, Kings Langley, WD4 8LZ, United Kingdom. \protect\\
E-mail: enrique.delucas@imgtec.com

\IEEEcompsocthanksitem J.L. Arag\'{o}n is with the Computer Engineering Dept., University of Murcia,
Campus de Espinardo, 30100 Murcia, Spain. \protect\\
E-mail: jlaragon@ditec.um.es}

}

%
%

{Anglada \MakeLowercase{\textit{et al.}}: Dynamic Sampling Rate: Harnessing Frame Coherence in Graphics Applications for Energy-Efficient GPUs}
%



\IEEEtitleabstractindextext{%
\begin{abstract}
In real-time rendering, a 3D scene is modelled with meshes of triangles that the GPU projects to the screen. They are discretized by sampling each triangle at regular space intervals to generate fragments which are then added texture and lighting effects by a shader program. Realistic scenes require detailed geometric models, complex shaders, high-resolution displays and high screen refreshing rates, which all come at a great compute time and energy cost.
This cost is often dominated by the fragment shader, which runs for each sampled fragment. Conventional GPUs sample the triangles once per pixel, however, there are many screen regions containing low variation that produce identical fragments and could be sampled at lower than pixel-rate with no loss in quality. Additionally, as temporal frame coherence makes consecutive frames very similar, such variations are usually maintained from frame to frame. This work proposes Dynamic Sampling Rate (DSR), a novel hardware mechanism to reduce redundancy and improve the energy efficiency in graphics applications. DSR analyzes the spatial frequencies of the scene once it has been rendered. Then, it leverages the temporal coherence in consecutive frames to decide, for each region of the screen, the lowest sampling rate to employ in the next frame that maintains image quality.
We evaluate the performance of a state-of-the-art mobile GPU architecture extended with DSR for a wide variety of applications. Experimental results show that DSR is able to remove most of the redundancy inherent in the color computations at fragment granularity, which brings average speedups of 1.68x and energy savings of 40\%.
\end{abstract}

}

\maketitle

\IEEEdisplaynontitleabstractindextext

%
\IEEEpeerreviewmaketitle

\IEEEraisesectionheading{\section{Introduction}\label{sec:introduction}}
The growing computing capabilities of today's mobile devices allow for real-time rendering of complex 3D scenes. Supporting  users' demands for ever richer applications clashes with the battery-operated nature of those devices, which makes energy efficiency paramount. In particular, previous studies have described the GPU as the most energy-demanding component of mobile SoCs for graphics workloads~\cite{patil2015characterization, anandtechGPU, 8675248}.

In real-time graphics, the geometry of objects is modelled with a set of vertices, which the GPU processes, assembles into flat polygons (normally triangles), and projects to the screen plane. Triangles are then discretized into arrays of pixels by sampling their surfaces at regular intervals to produce a fragment (i.e., a set of attributes such as color, depth, normal, etc.) for each sampled location, often just once per pixel. Fragments are then conveniently textured, shaded and blended to obtain a final color value per pixel.

On the one hand, if parts of the scene contain high spatial frequencies, sampling triangles at a low rate may be insufficient to capture fine details and would cause aliasing effects, such as jagged edges or flickering. Supersampling is an approach that relieves these artifacts by sampling at higher rates and combining the results into a single color. However, it involves large energy and performance costs since it increases the number of generated fragments \cite{maule2012transparency}, and it is well documented that fragment processing is the most energy consuming stage of the graphics pipeline due to the amount of computations and memory accesses required \cite{shebanow2013evolution,pool2012energy,enriqueTesis}. For that reason, supersampling is rarely applied nowadays in real-time rendering applications, especially in mobile GPUs, and scenes are usually rendered at one color sample per pixel.

\begin{figure*}[!htb]
    \centering
\includegraphics[height=95mm,width=.6\textwidth]{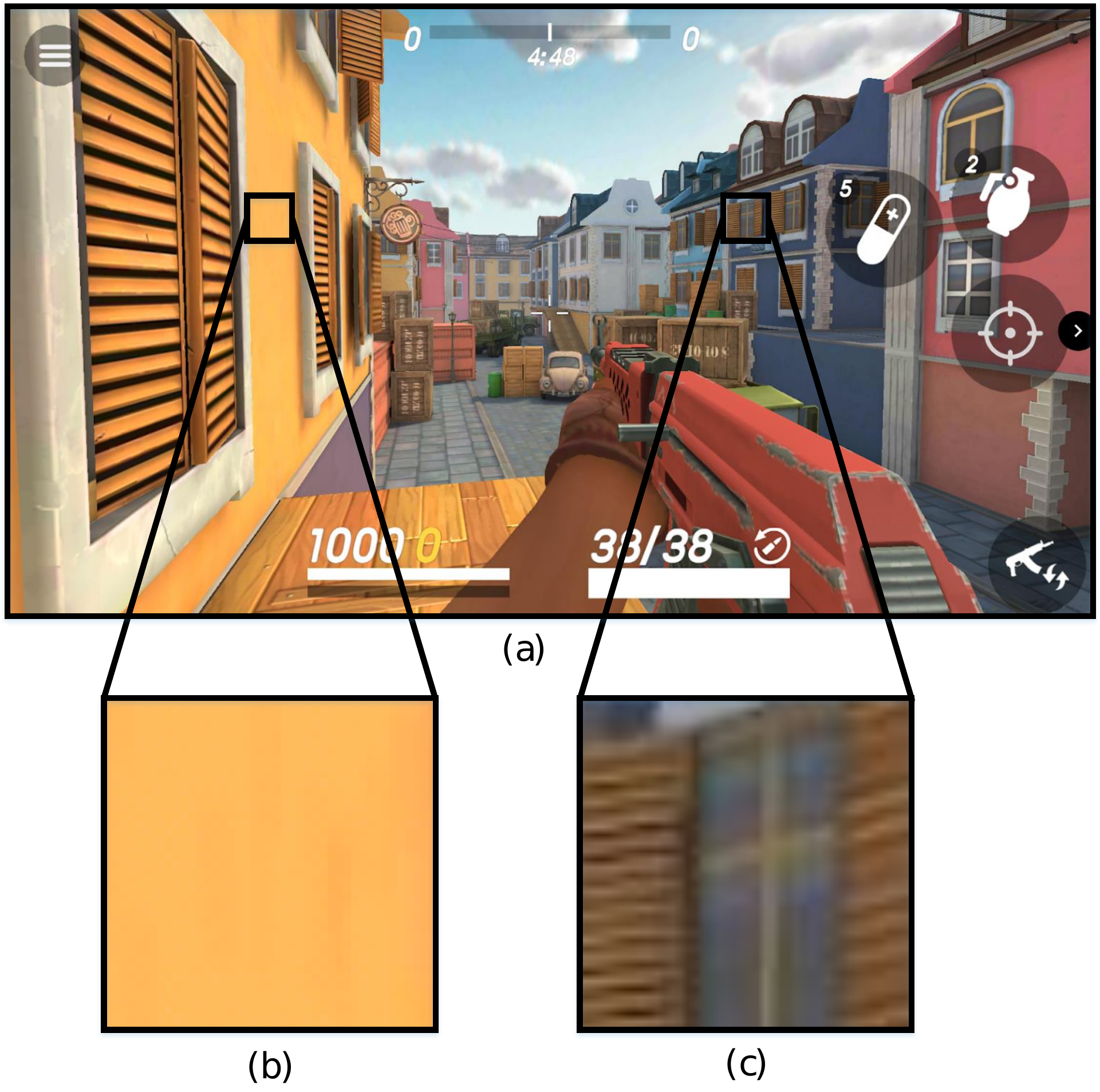}
    \caption{Difference in level of detail across a frame. a) Frame of the game \textit{Guns of Boom}. b) Region with low level of detail. c) Region with significant level of detail.}
    \label{fig:motivation}
\end{figure*}

\begin{figure}[!htb]
    \centering
\includegraphics[width=.45\textwidth]{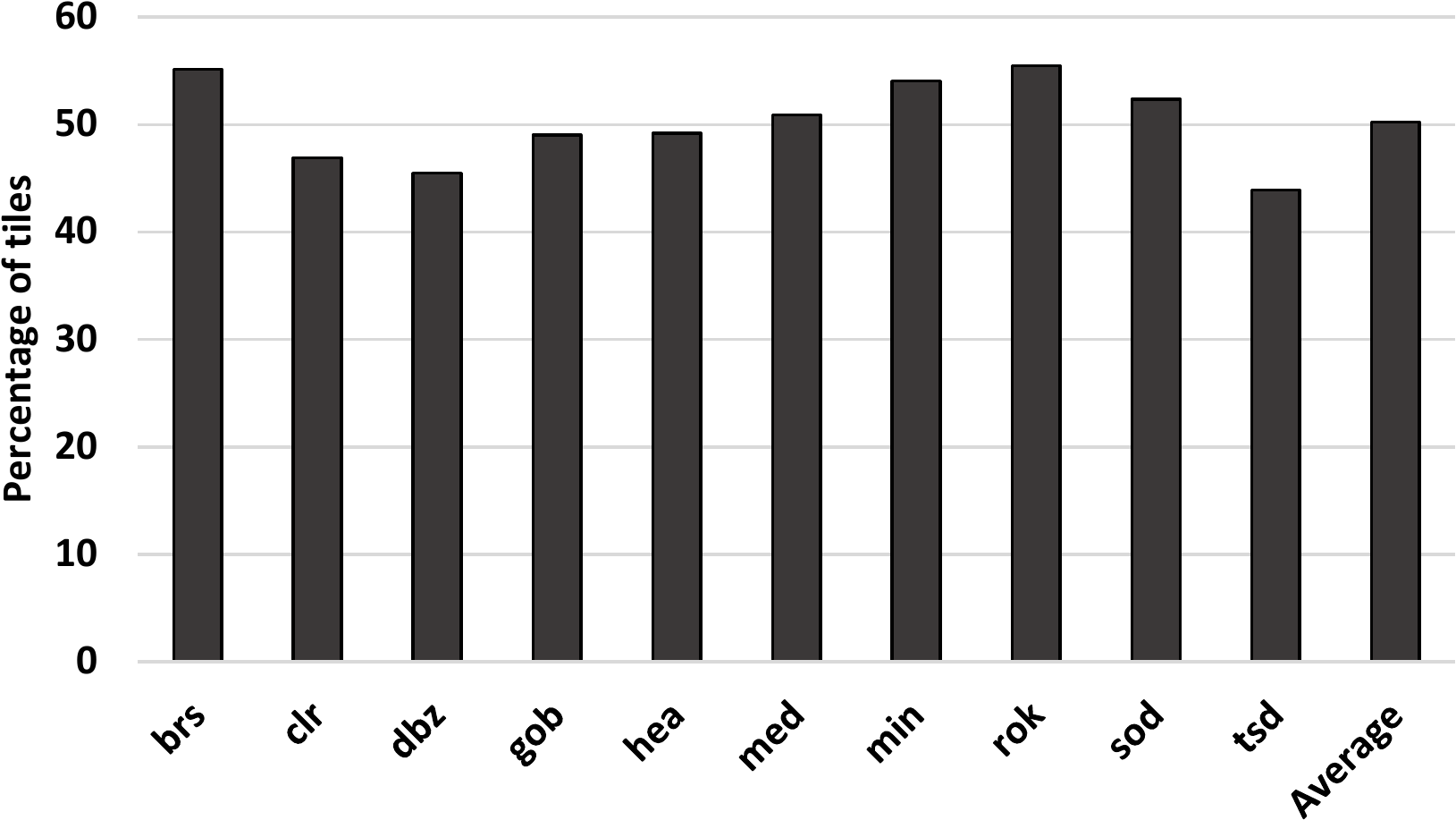}
    \caption{Number of 16x16 tiles that can be sampled at a rate lower than one sample per pixel without generating per-tile visible artifacts. Section \ref{sec:parameters} describes the methodology employed for this categorization.}
    \label{fig:potential}
\end{figure}

On the other hand, sampling all triangles at the same rate is not efficient because not all parts of the screen require the same sampling rate \cite{vaidyanathan2014coarse}. Figure \ref{fig:motivation} illustrates this phenomenon by comparing two different regions of the screen in a given frame: while \ref{fig:motivation}c contains significant level of detail, \ref{fig:motivation}b is homogeneous with a single color and, therefore, does not require per-pixel sampling. We have quantified, for a variety of mobile graphics applications, the number of 16x16-pixel regions of the screen that do not contain enough level of detail for them to require one sample per pixel. Figure \ref{fig:potential} shows that on average almost half of the screen can be processed at a lower sampling rate without affecting image quality. Properly identifying and removing the large amount of resources devoted to these unnecessary computations can lead to a substantial reduction in energy consumption.

Based on the above observations, this paper presents Dynamic Sampling Rate (DSR): a hardware mechanism that dynamically finds and applies, for each part of the scene, the optimal sampling rate, i.e., the lowest sampling rate that does not cause visible artifacts in the rendered image. DSR is designed for Tile-Based Rendering (TBR) architectures \cite{akenine2008graphics}, a common pipeline organization in mobile devices that divides the screen into rectangular sections -tiles- and renders them in succession, allowing the storage of temporary values in on-chip buffers to avoid their corresponding accesses to main memory. After the rendering of a tile to the on-chip color buffer finishes, DSR computes the Discrete Cosine Transform of the resultant tile image, analyzes the spatial frequencies present in it and decides, based on a simple heuristic, the best sampling rate for the tile: whether it could have been sampled at a lower rate without sacrificing image quality, whether it contains enough detail that the sampling rate needs to be increased or whether the sampling rate is already optimal. The estimated-best rates for all the tiles are then stored in a small on-chip Lookup Table and are queried during the following frame, before the rendering of each tile. DSR estimates the best sampling rate for a tile in one frame and applies it in the following frame because it takes advantage of the well-known frame coherence property of graphics animations \cite{Hubschman1982}: visual smoothness is achieved by producing a (quick) succession of frames where a large subset of tiles are very similar between two consecutive frames.

DSR addresses the shortcomings of prior work on the area of sampling at coarser granularities to reduce the number of shader executions. The majority of approaches rely on heavy efforts from the programmer to specify which components of the scene contain less details such as certain lights \cite{he2014extending} or vertex attributes \cite{vaidyanathan2014coarse}. Conversely, DSR decides the sampling rates in a completely transparent manner to the programmer. Unlike approaches in which a static sampling rate is set for parts of the scene (such as particular regions of the screen \cite{nvidiavrs} or fragments in the boundary between triangles \cite{sathe2015pixel}), DSR dynamically and continuously adapts the sampling rate of the entire scene at tile granularity to closely track image changes. Additionally, to the best of our knowledge, DSR is the first work to take advantage of frame coherence for this purpose. Although frame coherence has been proposed to skip some shader executions in the so-called Checkerboard rendering \cite{mcferroncheckerboard}, such scheme is lossy and may affect image quality. DSR, on the other hand, only reduces the sampling rate in tiles that do not contain high spatial frequencies and, therefore, DSR does not produce visual artifacts. Previous approaches that dynamically change the sampling rate must compute their estimates in the middle of the pipeline execution \cite{stengel2016adaptive}, whereas DSR is architected to not introduce any time overhead by overlapping the frequency analysis of one tile with the rendering of the next one.

To summarize, the main contributions of this paper are:

\begin{itemize}

\item A new hardware technique, completely transparent to the programmer, that estimates the lowest possible sampling rate to which each tile may be rendered without producing visual artifacts and applies it during the following frame by taking advantage of frame coherence.

\item A dynamic mechanism based on real-time analysis of the spatial frequencies that continuously adapts the per-tile sampling rate to track the image changes that occur over time.

\item A comprehensive architectural description of the frequency analysis unit and how it is integrated within the graphics pipeline in a way that it causes no timing overheads.

\item An implementation and evaluation that shows that Dynamic Sampling Rate reduces the shading activity by 66\% and memory accesses by 28\%, yielding speedups of 1.68x and an overall energy reduction of 40\% on average.

\end{itemize}

The rest of the paper is organized as follows. Section \ref{sec:bk} reviews the common pipeline organization of mobile GPUs and the Discrete Cosine Transform as a mean to mapping an image into the frequency domain. Section \ref{sec:dsr} describes the design of DSR. Section \ref{sec:parameters} presents the approach used to set the DSR's parameters. Section \ref{sec:implementation} illustrates the implementation of DSR and the changes required to the baseline GPU. Section \ref{sec:experimental} presents the experimental framework and Section \ref{sec:results} quantifies the benefits of applying DSR. Section \ref{sec:rw} outlines some related work and Section \ref{sec:conclusions} sums up the main conclusions of this work.

\section{Background} \label{sec:bk}
\subsection{Tile-Based Rendering}
\begin{figure}[!htb]
    \centering
\includegraphics[width=.47\textwidth]{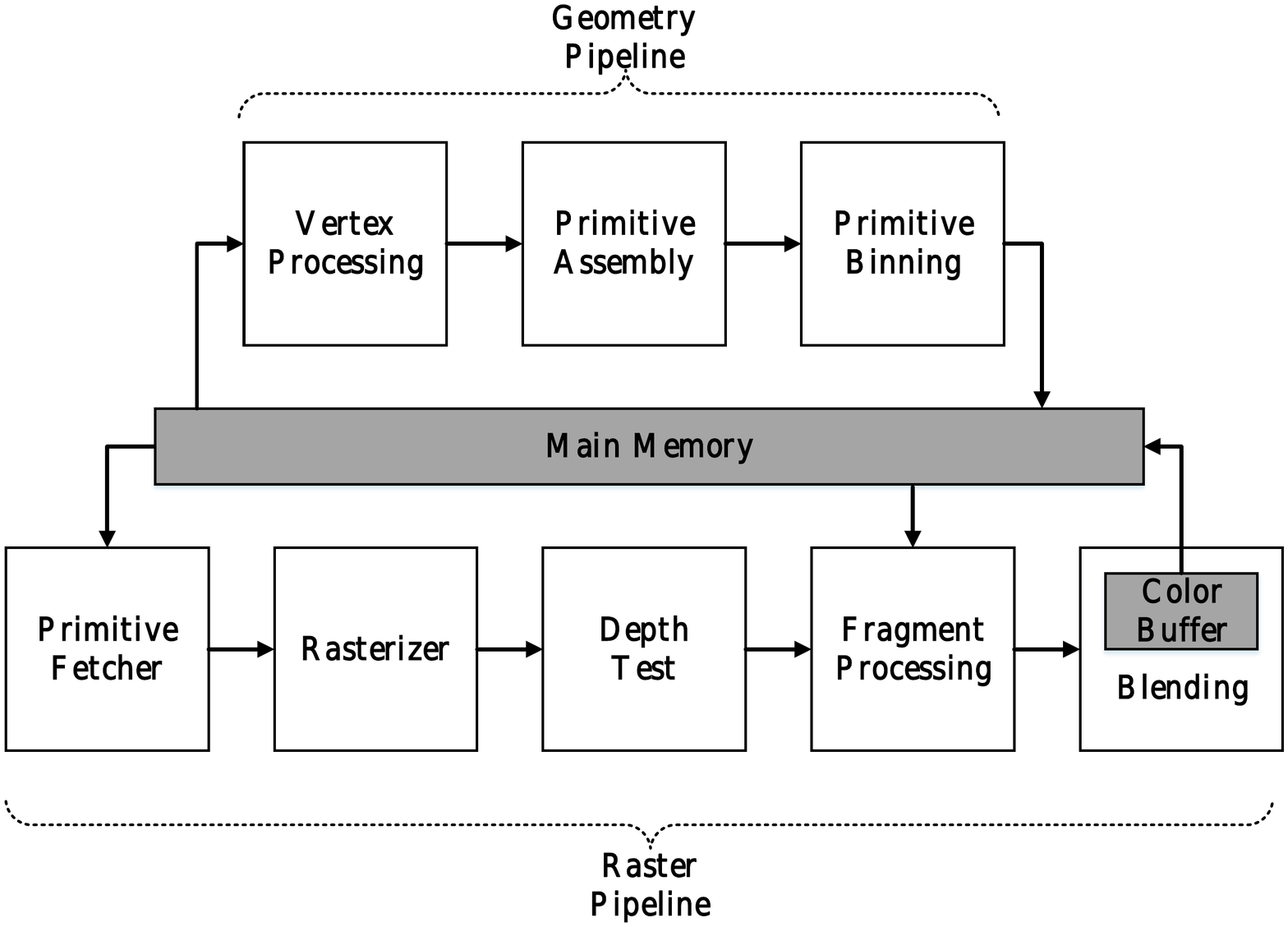}
    \caption{Tile-Based Rendering GPU.}
    \label{fig:basepipe}
\end{figure}

In a Tile-Based Rendering GPU, the image is rendered in two decoupled steps, as shown in Figure \ref{fig:basepipe}. First, the Geometry Pipeline processes the vertices of the three-dimensional models, transforming and assembling them into triangles. Triangles are then binned into the tiles that they overlap. Once the whole geometry has been processed and stored, the Raster Pipeline executes the second step, one tile at a time. By working on tiles, all temporary color values can be stored in the on-chip Color Buffer, which is only flushed to main memory once all triangles of the tile have been processed. The execution in the Raster Pipeline starts by fetching the primitives from memory and dispatching them to the rasterizer. Then, the rasterizer samples the surface of triangles at regular space intervals, generating \textit{Fragments}: points inside the triangle with interpolated information at each point. Adjacent fragments are then arranged into groups of four called \textit{Quads} and are sent to the Fragment Shaders which compute their colors by executing user-defined programs in lockstep mode for the four fragments. The color of each pixel in the tile is held in the Color Buffer, and it is obtained from one fragment, by blending multiple adjacent fragments, or by replicating the same fragment into multiple adjacent pixels, depending on whether the sampling rate is equal, higher or lower than one per pixel, respectively. 

\label{sec:background}
\subsection{Frequency Analysis}\label{sec:freqAnalysis}
The typical sampling rate of a scene is one sample per pixel. According to the Nyquist Sampling Theorem \cite{nyquist1928certain}, such a sampling rate allows capturing changes in the image every two pixels or more, i.e., with a frequency smaller or equal than two pixels. As shown in Figure \ref{fig:motivation}, not all regions of the screen contain high frequencies, or changes in the image in a short space. Therefore, a (much) lower sampling rate may be enough to represent the original signals. We propose to analyze the frequencies of each rendered tile to decide the sampling rate to be applied to it in the following frame.

\begin{figure}[!htb]
    \centering
\includegraphics[]{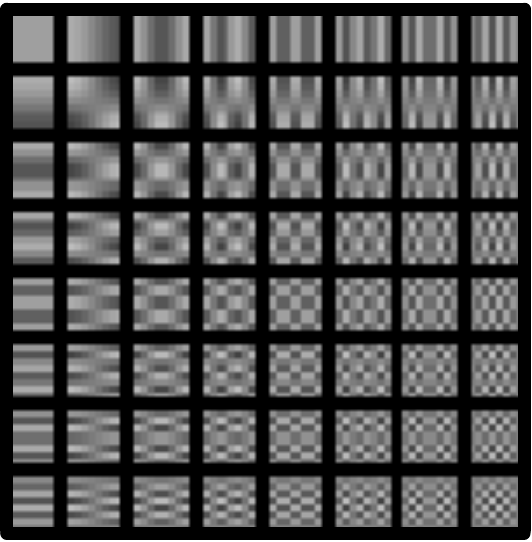}
    \caption{DCT basis functions for N=8 pixels.}
    \label{fig:basis}
\end{figure}

A well known mechanism to obtain the frequency components of an image is the Discrete Cosine Transform (DCT) \cite{ahmed1974discrete}. As a Fourier-related transform, the DCT maps a function (an image) from the spatial domain to a set of coefficients of basis functions localized in the frequency spectrum. Those basis functions correspond to sinusoids of a certain frequency and are visually represented in Figure \ref{fig:basis}. It can be seen that as either the $x$ or $y$ axis increase, the basis function is a sinusoid with higher variation rate, i.e., with higher frequency. Applying a 2D DCT to a block of NxN pixel colors results in a NxN matrix of values, the coefficients of the linear combination of basis functions which represent the original image in the frequency domain.  The coefficient present in each element of the matrix indicates how much of that particular frequency is found in the original image.

The 2D DCT has several characteristics that make it an ideal choice for the type of real-time frequency analysis we require to find the optimal sampling rate for a tile:
\begin{itemize}
    \item It assumes an even symmetry of the function: by construction, the image is mirrored in all its borders, which avoids artificial high frequency components that other transforms introduce by only considering a NxN pixel subset of the image.
    \item It has very high energy compaction, which means that the great majority of frequency information is summarized in the upper-left region of the result matrix. This allows us to make sampling rate decisions only considering a subset of the NxN coefficients.
    \item It has a low complexity cost in comparison with other transforms as only cosines are computed.
\end{itemize} 

Additionally, the 2D DCT is a separable function, which allows for the linear computation of all the elements in one dimension followed by the linear computation of all the elements in the second dimension. These characteristics allow us to implement a fast and energy efficient hardware unit to analyze the frequency components of a tile, explained in more detail in Section~\ref{sec:implementation}.

\section{Dynamic Sampling Rate}\label{sec:dsr}
This section describes how the 2D DCT is used to estimate the optimal sampling rate for a tile, i.e., the lowest sampling rate that does not introduce visible artifacts in the overall frame, and how to dynamically adapt it to image changes over time.

When the rendering process of the tile finishes, the Color Buffer contains the final color for all the pixels of the tile. We propose to add a small hardware unit that takes these colors as inputs to compute the 2D DCT and analyzes the resulting matrix of coefficients to determine if the current sampling rate for the tile is optimal. All the DCT coefficients are first aggregated into a single value that summarizes the amount of high-frequency information of the tile. Although a plethora of metrics exist, we empirically determined that the maximum absolute value among the coefficients corresponding to high-frequency diagonals suffices, and we will refer to it as \textit{MaxC} (we term here diagonal $k$ as the set of all elements of the matrix whose row index plus column index is equal to $k$: for instance, diagonal 3 consists of elements $(0,3), (1,2), (2,1)$ and $(3,0)$). The rationale under this choice is that, intuitively, we are more interested in knowing if the largest high-frequency component is big enough to justify a high sampling rate rather than considering the effect of multiple high-frequency components combined. The low-frequency components of the matrix are not taken into account in the computation of \textit{MaxC}. 

\begin{figure}[!htb]
    \centering
\includegraphics[width=0.47\textwidth]{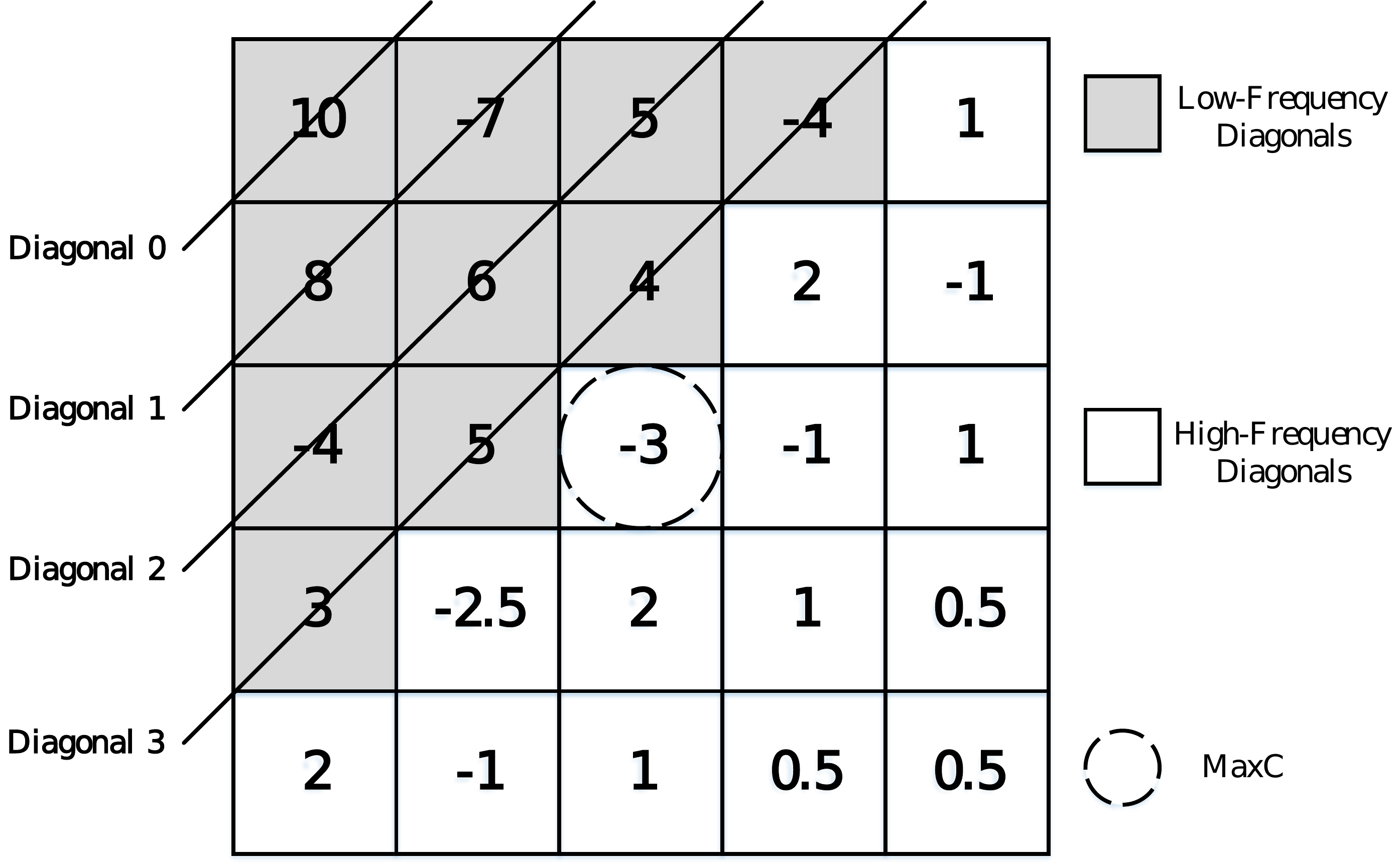}
    \caption{\textit{MaxC} determination example.}
    \label{fig:maxc}
\end{figure}

Figure \ref{fig:maxc} illustrates the determination of \textit{MaxC} in a 5x5 coefficient matrix in which we consider diagonals 0 through 3 as low-frequency diagonals. As shown, \textit{MaxC} is 3, since it is the highest absolute value among all the high-frequency diagonals. Although larger values appear in the low-frequency diagonals, they are ignored.

Then, a simple test is conducted to decide the new sampling rate for the tile: \textit{MaxC} is compared against two different thresholds.

\begin{itemize}
    \item The first threshold, which we label \textit{Reduce Threshold} ($T_{R}$), represents the maximum frequency a tile can contain for it to be sampled at a rate lower than the current one. If \textit{MaxC} is lower than the Reduce Threshold, the sampling rate for the tile is reduced. 
    \item The second threshold, which we label \textit{Increase Threshold} ($T_{I}$) represents the maximum frequency a tile can contain for it to be sampled at the current rate. If \textit{MaxC} is greater than the Increase Threshold, the sampling rate for the tile is increased.
\end{itemize}

In the case that \textit{MaxC} is neither lower than the Reduce Threshold nor greater than the Increase Threshold, the sampling rate for the tile does not change. The new sampling rate for each tile is stored and used to process it in the next frame. The scene is, therefore, not sampled uniformly neither in space nor time: each tile is rasterized with an independent sampling rate and it may be modified across frames to adapt to image changes. 

\begin{figure}
    \centering
\includegraphics[width=0.47\textwidth]{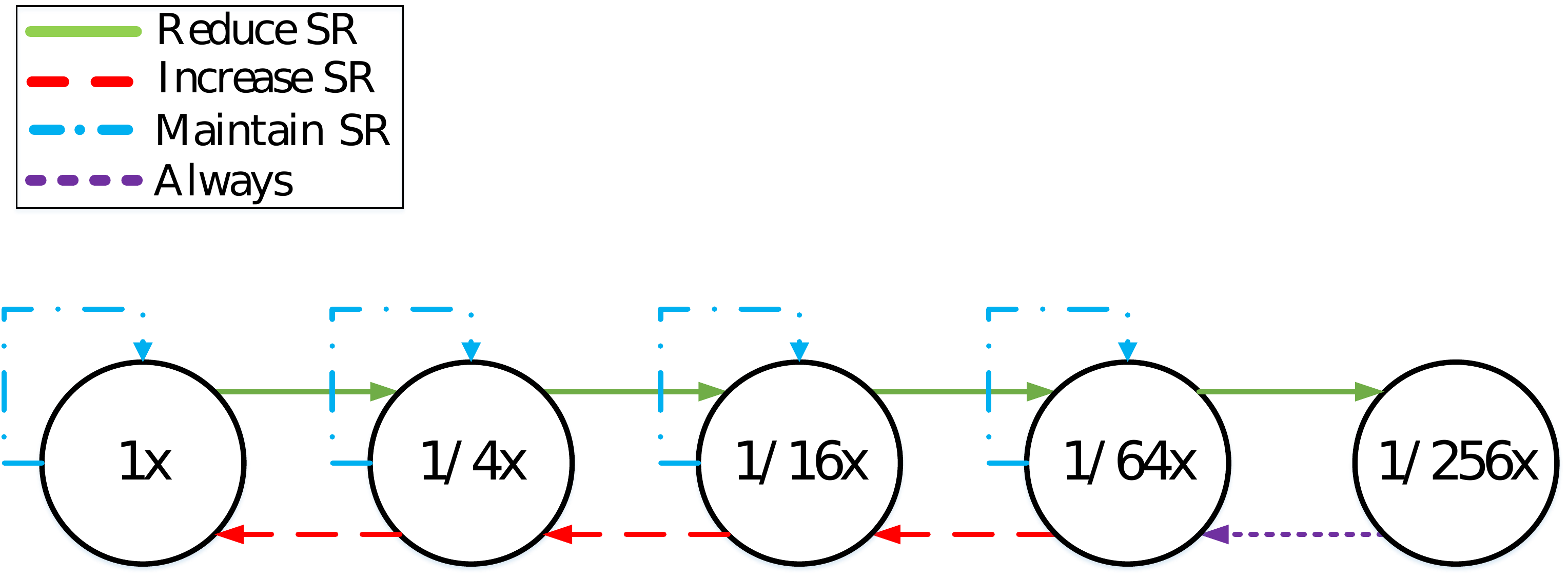}
    \caption{Dynamic Sampling Rate Finite-State Machine.}
    \label{fig:fsm}
\end{figure}

\begin{figure}[!htb]
    \centering
\includegraphics[width=.45\textwidth]{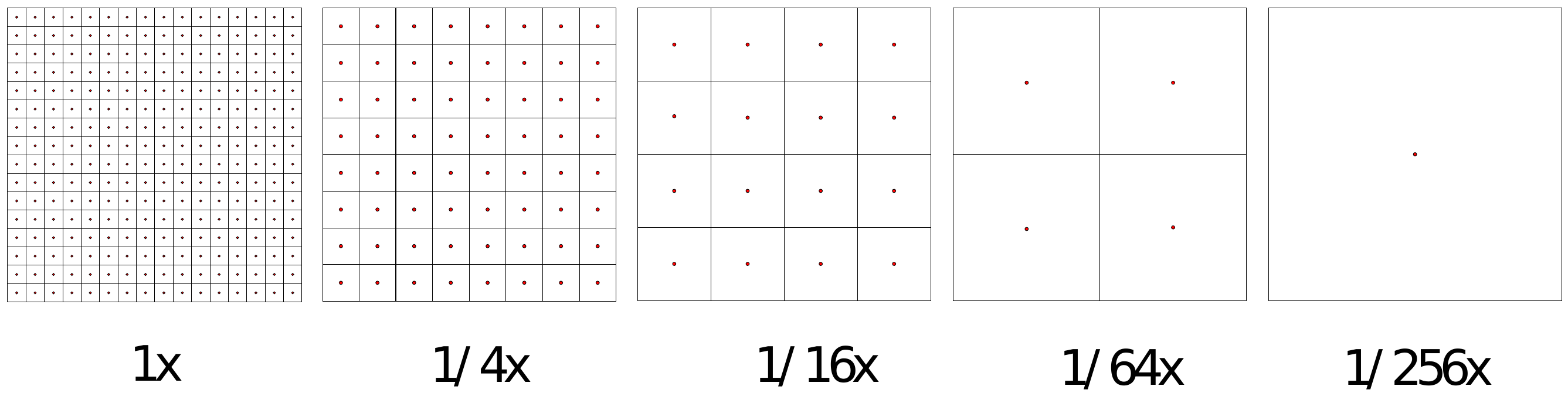}
    \caption{The five sampling rates considered in our experiments, from 1x (left) to 1/256x (right).}
    \label{fig:sample_rates}
\end{figure}

Figure \ref{fig:fsm} shows the FSM that manages the dynamic sampling rate determination. We consider five different sampling rates: sampling at the center of every pixel (baseline sampling rate) and sampling at the center of every square block of 4, 16, 64 or 256 pixels (as shown in Figure \ref{fig:sample_rates}). We will refer to these sampling rates as 1x, 1/4x, 1/16x, 1/64x and 1/256x, respectively. These sampling rates are motivated by the baseline GPU architecture employed in this work, which utilizes tiles of 16x16 pixels. Each state in the FSM corresponds to halving the previous sample rate in both $X$ and $Y$ dimensions, and the lowest state only generates one sample per tile. The transitions among states are controlled by the heuristic decision described above, based on a $<\!\!T,D\!\!>$ tuple that contains: the Thresholds ($T$) to which \textit{MaxC} is compared to, and the number of low-frequency matrix Diagonals that are ignored ($D$) for its computation. We label as $<\!\!T_{R}, D_{R}\!\!>$ the tuples for the Reduce transitions and as $<\!\!T_{I}, D_{I}\!\!>$ the tuples for the Increase transitions.

As images generated with lower sampling rates have fewer high-frequency components and different sampling rate requirements, each transition in the FSM has individual values for $<\!\!T_{R}, D_{R}\!\!>$ and for $<\!\!T_{I}, D_{I}\!\!>$. Apparently, the FSM has 4 Increase and 4 Decrease transitions. However, at 1/256x rate, fragments are sampled just once and the resulting tile contains a single plain color. As there is no spatial frequency in it, the heuristic cannot make decisions based on the coefficient matrix. Our FSM conservatively forces the 1/256x state to always transition back to 1/64x. Consequently, we must set parameter values for 3 Increase and 4 Reduce transitions in the FSM. Adequate values for these parameters have been empirically determined through extensive experiments with the objective to reduce GPU activity (samples) while keeping the original image quality. Section \ref{sec:parameters} describes the methodology followed to find such optimal $<\!\!T_{R}, D_{R}\!\!>$, $<\!\!T_{I}, D_{I}\!\!>$ values for each sampling rate.

Although in this work we consider sampling once per fragment to be the highest sampling rate, DSR can easily be integrated in GPUs that allow higher sampling rates, such as the ones implementing Supersampling Antialiasing (SSAA). The maximum sampling rate that those GPUs provide will be used as DSR's baseline sampling rate. DSR will then selectively apply SSAA by dynamically determining which regions of the screen require it. DSR can also be combined with the widely spread Multisampling Antialiasing (MSAA) approach \cite{akeley1993reality}. With MSAA, depth is sampled at several points per pixel (usually 4), while the fragment shader is only computed at the pixel center. All the samples that pass the depth test receive the output of the shader. As DSR only changes the frequency in which shaders are executed, DSR can also be applied on systems that use MSAA.

\section{Parameter Selection}\label{sec:parameters}
This section describes the empirical methodology to find the best values for DSR parameters such that frames are rendered at the lowest possible average sample rate (ASR) without producing any visible error.

As depicted in Algorithm \ref{alg:heuristic}, we perform an exhaustive parameter exploration. For each parameter combination under test we render all frames, adjusting the sample rate of each tile according to the output of the heuristic. During the search, any combination that produces even a single erroneous frame is directly discarded. Otherwise, the achieved ASR across all frames is computed for that combination. Eventually, we choose the parameter combination that produces the lowest ASR.

Frame errors are computed by comparing the image quality of the produced frames with respect to the frame rendered at baseline sampling rate using the Mean Structural Similarity Index (MSSIM \cite{wang2004image}), a widely adopted, perceptually-based quality metric that estimates the visual impact of changes in image luminance and contrast caused by compression distortions. The MSSIM has been shown to outperform other similarity metrics that just measure differences in pixel color, such as PSNR and MSE, in terms of quality \cite{gao2009image,ma2010new} as it correlates better with the perception of the human visual system. A frame error occurs whenever the obtained MSSIM is lower than 95, as it is the point at which defects can be discerned by human beings \cite{flynn2013image}.

\begin{algorithm}
\caption{Basic parameter search}
\hspace*{\algorithmicindent} \textbf{Input:}   \\
\hspace*{\algorithmicindent} \textbf{Output: Values for the $<T_{I},D_{I}>$ and \newline $<T_{R},D_{R}>$ tuples that produce the lowest ASR} 
\label{alg:heuristic}
\begin{algorithmic}[1]
    \For{\textbf{each} parameter combination}  
        \For{\textbf{each} frame}
            \For{\textbf{each} tile}
                \State $DCT = $ compute\_dct($tile, frame$)
                \If{$MaxC(D_{R}, DCT) < T_{R}$} 
                    \State  $next = SR[tile, frame]-1$ \Comment{Reduce}
                \ElsIf{$MaxC(D_{I}, DCT) \geq T_{I}$} 
                    \State $next = SR[tile, frame]+1$ \Comment{Increase}
                \Else 
                    \State $next = SR[tile, frame]$ \Comment{Stay}
                \EndIf
                \State $SR[tile, frame+1] = next$
            \EndFor
            \If{contains\_errors($frame$)}
                \State discard parameter combination
            \EndIf
        \EndFor
        \State compute\_ASR($parameter\_ combination, SR$)
    \EndFor
\end{algorithmic}
\end{algorithm}

Each parameter combination contains a set of 14 different parameters (four $<\!\!T_{R},D_{R}\!\!>$ pairs for the Reduce transitions and three $<\!\!T_{I}, D_{I}\!\!>$ pairs for the Increase transitions). Even considering just a few values for each parameter (say $n$), the sheer amount of combinations to consider ($n^{14}$) makes an exhaustive exploration unfeasible. We adopt instead a divide and conquer approach in which we first only focus on finding the best parameters for the Increase transitions. Next, those values are used and kept constant in Algorithm \ref{alg:heuristic} to find the best parameters for the Reduce transitions. By splitting the parameter search into two steps, we substantially limit the number of combinations to explore and we can execute an exhaustive search.

Note however that during the first step we cannot apply Algorithm \ref{alg:heuristic}: without values set for the $<\!\!T_{R},D_{R}\!\!>$ pairs, the procedure lacks a mechanism to dynamically reduce the sample rates and the FSM never reaches the lowest states. Consequently, we must provide an alternative sampling rate reduction mechanism for this first step. Such mechanism must produce tiles at low enough sampling rates that Increase transitions are required to prevent errors due to undersampling. Otherwise (if Increase decisions were never required) we would not test the capabilities of the parameter combinations to produce a low ASR while not producing frame errors.

To build an effective reduction mechanism we first conduct a simple preliminary experiment that finds near-optimal sample rates for each of the tiles in all frames. Those values will act as references and will stay constant during the exploration of the Increase parameters. The reduction mechanism consists in always choosing the lowest sampling rate between the reference value and the outcome of the heuristic (which either increases the sampling rate or keeps it the same).

This preliminary experiment first generates the images of all tiles in all frames at all five sampling rates. It then sequentially analyzes tile by tile the five alternatives and selects the lowest one that does not produce visible errors compared with the same tile at baseline sampling. We term these sample rates \textit{Local Minimum}, because image discrepancies are not analyzed at full frame level but just at tile level. As such, they may not be the optimal sample rates (optimal values may be lower when discrepancies are analyzed at frame level) but they are low enough to be used as a reference in our reduction mechanism.

Algorithm \ref{alg:locmin} shows the procedure to find the best parameters for the Increase transitions (the first step). Akin to Algorithm \ref{alg:heuristic}, for each tile it computes the DCT and decides whether or not to increase the current sampling rate according to the $<\!\!T_{I},D_{I}\!\!>$ parameters under test (Lines \ref{line:if}-\ref{line:endif}). However, unlike Algorithm \ref{alg:heuristic}, it next considers overriding that decision by choosing instead the stored Local Minimum sample rate for the next frame (Line \ref{line:min}) in case that it is lower. As the algorithm can select a sampling rate lower than the Local Minimum, the found parameters gravitate towards the optimal sampling rates.

\begin{algorithm}
\caption{LocMin parameter search}
\hspace*{\algorithmicindent} \textbf{Input:LocalMinimum, $<T_{R},D_{R}>$ }   \\
\hspace*{\algorithmicindent} \textbf{Output:Values for the $<T_{I},D_{I}>$ tuples that produce the lowest ASR} 
\label{alg:locmin}
\begin{algorithmic}[1]
    \For{\textbf{each} parameter combination}  
        \For{\textbf{each} frame}
            \For{\textbf{each} tile}
                \State $DCT = $ compute\_dct($tile, frame$)
                \If{$MaxC(D_{I}, DCT) \geq T_{I}$} \label{line:if}
                \State $next = SR[tile, frame]+1$ \Comment{Increase} 
                \Else 
                    \State $next = SR[tile, frame]$ \Comment{Stay}
                \EndIf \label{line:endif}
                \State $locmin = LocalMinimum[tile, frame+1]$
                \State $SR[tile, frame+1] = min(next, locmin)$ \label{line:min}
            \EndFor
            \If{contains\_errors($frame$)}
                \State discard parameter combination
            \EndIf
        \EndFor
        \State compute\_ASR($parameter\_ combination, SR$)
    \EndFor
\end{algorithmic}
\end{algorithm}

\section{Implementation}\label{sec:implementation}
This section describes the combinational logic and memory structures required to implement Dynamic Sampling Rate, and how the frequency analysis and sample rate determination are integrated within the Raster Pipeline.

\subsection{Pipeline Integration}

The Dynamic Sample Rate technique uses a FSM (see Figure \ref{fig:fsm}) to dynamically determine the sampling rate of each tile based on its current state and its \textit{MaxC}. It requires a new hardware structure called \textit{Sampling Rate Table} (SRT), with one entry per tile, that holds the state of each tile in a frame. Since the FSM has 5 different states, a state can be represented with 3 bits. Consequently, for a frame resolution of 1080x1920 pixels (as modelled in our experiments) there are 8100 tiles and the storage overhead of the SRT is 2.96 KB.

\begin{figure}[!htb]
    \centering
\includegraphics[width=.48\textwidth]{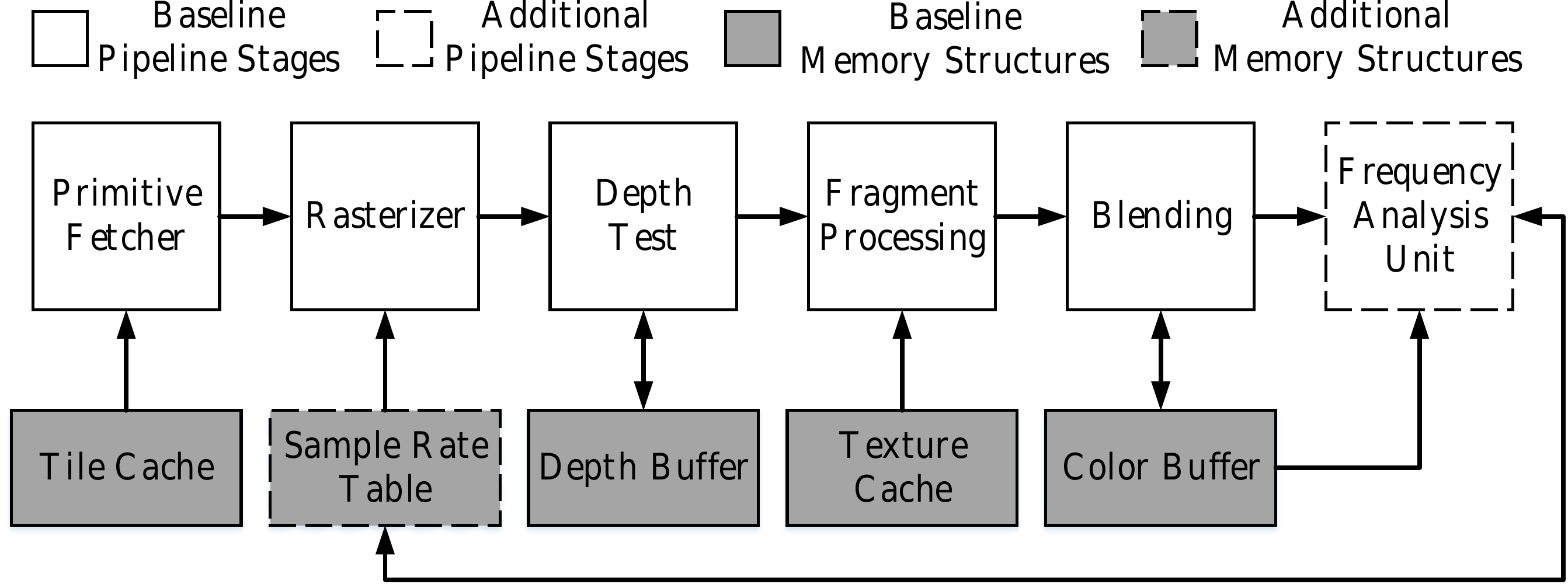}
    \caption{Raster Pipeline with DSR.}
    \label{fig:newpipe}
\end{figure}

Other than the SRT, Dynamic Sampling Rate requires very minor modifications to the pipeline, as shown in Figure \ref{fig:newpipe}. Tiles are scheduled and primitives are fetched in the same way as in the baseline because the sampling rate only affects the discretization process. The Rasterizer still produces Quads (square groups of four adjacent fragments), so they can be depth tested, shaded and blended as in the baseline. The main difference is that the screen area covered by each fragment is bigger than a pixel when the sampling rate is lower than 1x. We refer to those fragments as \textit{Superfragments} and to a group of four Superfragments as \textit{Superquads}. Producing a superfragment at a sampling rate of 1/NxN only requires sampling at the center of a grid of NxN pixels.

Whenever a tile starts its processing, its state (hence the associated sampling rate) is fetched from the Sampling Rate Table. The Rasterizer generates Superfragments according to the stored state. The Depth and Color Buffers already have capacity to hold temporary values for the 256 pixels (16x16 pixel tiles) of the baseline resolution. Fragments within a Superfragment share depth and color, so, only one read/write operation in the Depth Buffer is executed when depth testing a Superfragment and only one read/write operation in the Color Buffer is executed when blending a Superfragment. This results in some entries of the Color Buffer not being initialized after a tile finishes its processing. In the last pipeline stage, the final color value of a Superfragment is upsampled by replicating a color to all pixels belonging to it. Afterwards, the contents of the Color Buffer are transferred to main memory and the DCT computation of the tile starts

\subsection{Frequency Analysis Unit} \label{sec:fau}

The 2D DCT is a separable function \cite{rao2014discrete}. This property allows to transform a NxN $Input$ image into the frequency domain by successively applying 1D transforms, first along the rows and then along the columns (or vice-versa). By considering separability, the well-known 2D-DCT formula can be rearranged as shown in Equation \ref{eq:dct2dseparable}:

\begin{equation}
\label{eq:dct2dseparable}
\resizebox{.9\hsize}{!}{
 $DCT(p,q) = \alpha(p)\alpha(q)\sum_{m=0}^{N-1}cos\frac{(2m+1)\pi p}{2N} \sum_{n=0}^{N-1} Input_{mn}cos\frac{(2n+1)\pi q}{2N}$
}
\end{equation}

where $0 \leq p,q \leq N-1$ and the scale factors $\alpha$ are defined as:

\begin{equation}
\alpha_{p} = \alpha_{q}
     \begin{dcases*}
       \frac{1}{\sqrt{N}} &\quad\text{if p = 0 or q = 0}\\
       \sqrt{\frac{2}{N}} &\quad\text{otherwise.} \\ 
     \end{dcases*}
\end{equation}

The 2D-DCT formula is usually expressed in matrix notation as \cite{sihvo2005row}:

\begin{equation}
\label{eq:dctmatrix}
    DCT = K\;Input\;K^{T} = (K(K\;Input)^{T})^{T}
\end{equation}

where $K$ is the so-called \textit{Kernel Matrix}, that contains precomputed values for both the scale factors and the cosine functions in the form of:

\begin{equation}
K_{pq} =
     \begin{dcases*}
       \frac{1}{\sqrt{N}} &\quad\text{if } p = 0\\
       \sqrt{\frac{2}{N}}cos\frac{(2q+1)\pi p}{2N} &\quad\text{otherwise.} \\ 
     \end{dcases*}
\end{equation}

Our frequency analysis scheme uses the Synopsys's implementation of the 2D DCT transform from their DesignWare library~\cite{dwdct}. This module is based on the aforementioned Kernel Matrix precomputation and row-column decomposition. The computation of the 2D DCT shown in Equation \ref{eq:dctmatrix} is divided in two steps, decoupled by an auxiliary buffer ($Aux$) that holds temporary results. The first step computes the 1D-DCT of the rows ($Aux = (K\;Input)^{T}$) and the second step completes the 2D computation ($DCT = (K\;Aux)^{T}$). In the Synopsys implementation, a single buffer is used for storing both the temporary and final results. This buffer, which we name \textit{DCT Buffer}, is written by columns and read by rows to emulate the two transpositions. 

\begin{figure}[!htb]
    \centering
\includegraphics[width=.47\textwidth]{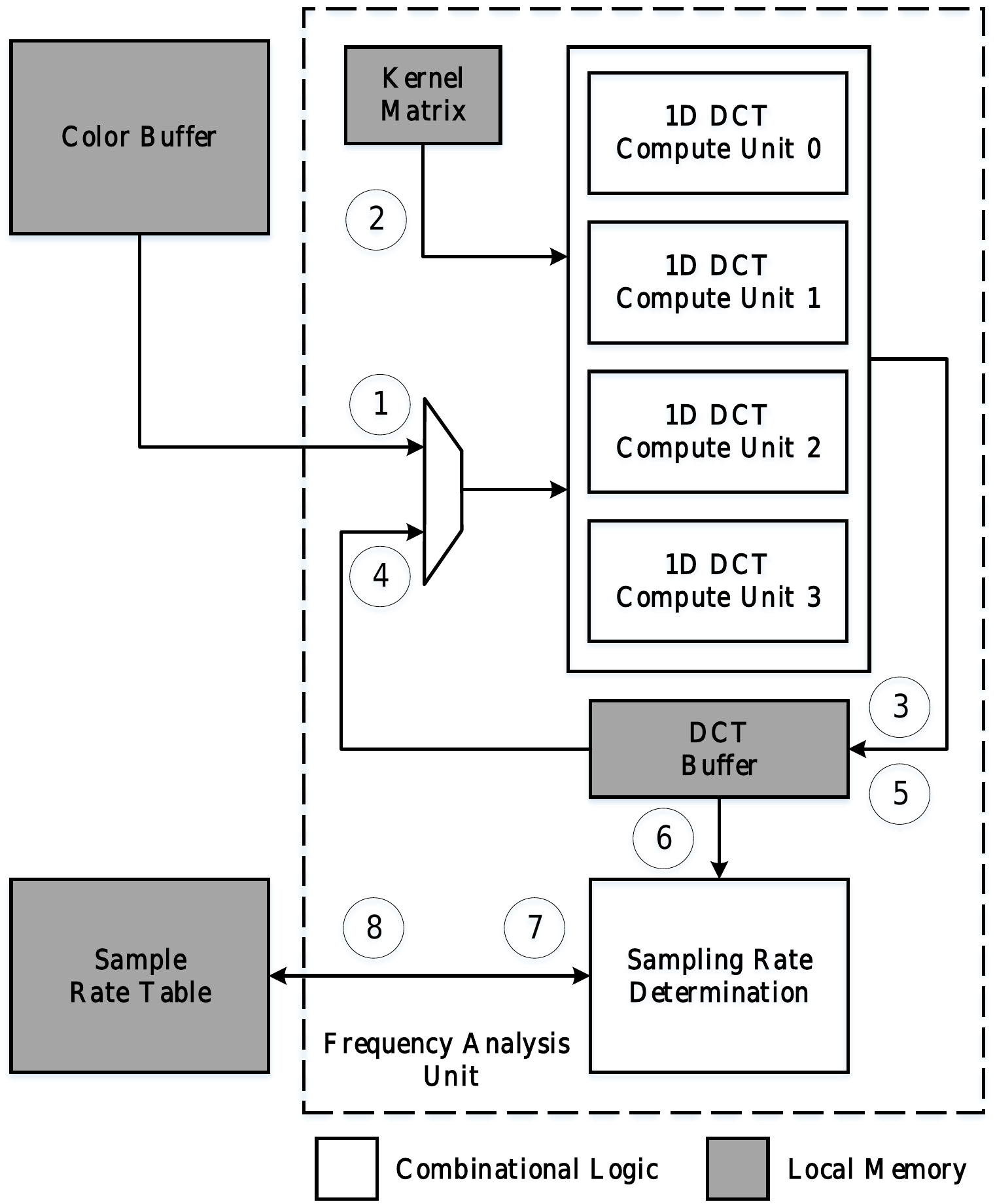}
    \caption{Frequency Analysis Unit overview.}
    \label{fig:fascheme}
\end{figure}

Figure \ref{fig:fascheme} shows a block diagram of the Frequency Analysis Unit and its dataflow: the input data is read from the Color Buffer \circled{1} and is multiplied by the Kernel Matrix \circled{2} using a series of compute units. Each unit computes the 1D-DCT of a row and stores the result in the DCT Buffer \circled{3}. Since the tiles in our modeled GPU are composed of 16x16 pixels and the frequency analysis unit contains 4 compute units, each unit sequentially processes 4 rows. Once the 16 rows have been processed, the second pass is performed: the temporary contents of the DCT Buffer \circled{4} are multiplied to the Kernel Matrix and stored back in the DCT Buffer \circled{5}, four columns at a time, until the final 2D DCT is computed.

The original Synopsys design operates sequentially in each row and column, as it does not have hardware to compute multiple 1D-DCTs in parallel. This implies significant time overheads to compute the entire 2D-DCT of 16x16 elements. We have slightly modified the design by replicating the compute units. Experimentally, we have determined that with 4 compute units the frequency analysis and sampling rate determination do not cause stalls in the pipeline and the energy and area overheads are minimal (results in Section \ref{sec:results}).

Once the DCT computation ends, the hardware \circled{6} estimates the best sampling rate for that tile using the scheme presented in Section \ref{sec:dsr}: it first uses the matrix of coefficients (ignoring the $D$ first diagonals) to compute \textit{MaxC}; then, following the FSM in Figure \ref{fig:fsm}, it decides the new tile state (hence a corresponding sampling rate), based on the current state \circled{7} and the comparison between \textit{MaxC} and the $T$ threshold. Finally, the new tile state is stored in the SRT \circled{8} to indicate the sampling rate to be used in the following frame.

\section{Experimental Framework}\label{sec:experimental}
\begin{table}[ht]
\centering
\caption{Benchmark suite.\strut}
\label{tab:benchmarks}
  \begin{adjustbox}{max width=0.47\textwidth}
\begin{tabular}{@{}lll@{}}
\toprule
\textbf{Benchmark}      & \textbf{Alias} & \textbf{Genre}  \\
\midrule
Brawl Stars   & brs            & Beat'em Up\\
Clash Royale   & clr            & Real-Time Strategy\\
Dragon Ball Z: Dokkan Battle   & dbz            & Board game, Puzzle\\
Guns of Boom   & gob           & First-Person Shooter\\
Hearthstone   & hea           & Collectible Card Game\\
Merge Dragons! & med & Puzzle\\
Minecraft & min & Sandbox\\
Rise of Kingdoms: Lost Crusade & rok & Real-time strategy\\
Sonic Dash & sod & Endless Runner\\
Toy Story Drop! & tsd & Puzzle\\

\bottomrule
\end{tabular}
\end{adjustbox}
\end{table}
The set of benchmarks employed in our experiments include ten unmodified commercial Android graphics applications that represent the current landscape of real-time rendering in mobile devices (see Table~\ref{tab:benchmarks}). The benchmark set consists of contemporary applications with tens of millions of downloads in Google Play \cite{googleplay} and includes a variety of workloads: from 2D applications with simple models (e.g., \textit{tsd}) to 3D applications with detailed scenes (e.g., \textit{gob}). The applications are also diverse in how the camera is placed and moved through the scene. The set includes benchmarks with the static-camera scenes (\textit{clr}, \textit{dbz}, \textit{hea}, \textit{tsd}) and simple scrolls (\textit{brs}, \textit{med}, \textit{sod}, \textit{rok}) that characterize mobile applications and also scenes with free-from and swift camera movements (\textit{gob}, \textit{min}).

The experimental framework used in this work is composed of three different stages:

\begin{enumerate}
    \item The benchmarks are first run on a smartphone equipped with an Adreno 530 GPU and a 5.15-inch, 1080p display. We use GAPID \cite{gapid} to obtain a file containing all the executed OpenGL commands of 100 frames of archetypal execution for each benchmark. GAPID is an open source debugging tool that captures the OpenGL API calls of an application to the graphics card driver. 

    \item The logged commands are then fed to the software-based back-end included in Gallium3D \cite{gallium}, which implements all the stages of the Graphics Pipeline and runs it on a CPU. We instrument the execution of the software renderer to obtain a complete instruction and memory trace of the application.
    
    \item The trace drives the execution of the cycle-accurate simulator of the TEAPOT toolset \cite{arnau2013teapot}, from which we obtain timing and energy results. The parameters used in our experiments are presented in Table \ref{tab:simulator} and model a TBR architecture resembling the ARM Mali-450 GPU \cite{mali450}. The simulator has been extended to include all the combinational logic and local memory structures required by DSR. Additionally, the Frequency Analysis Unit described in Section \ref{sec:fau} has been implemented in VHDL and synthesized to obtain its delay and power using the Synopsys Design Compiler, the modules of the DesignWare library and the 28/32nm technology library from Synopsys \cite{synopsys}.

\end{enumerate}

\begin{table}[ht!]
    \centering
    \caption{GPU Simulation Parameters. \strut}
    \label{tab:simulator}
     \begin{adjustbox}{max width=0.47\textwidth}
	\begin{tabular}{ l  p{5.5cm} }
	  \toprule	  
	  \multicolumn{2}{c}{\textbf{Baseline GPU Parameters}} \\
	  \bottomrule	  	  
	  \addlinespace[0.25em]
	  Tech Specs & 400 MHz, 1 V, 32 nm \\
	  Screen Resolution & 1080x1920 \\
	  Tile Size & 16x16 pixels \\
      \addlinespace[0.25em]
     \toprule
	  \multicolumn{2}{c}{\textbf{Main Memory}} \\
      \bottomrule
      \addlinespace[0.25em]
	  Latency &	50-100 cycles \\
	  Bandwidth	&	4 B/cycle (dual channel LPDDR3) \\
	  Size	&	1 GB  \\
	  \addlinespace[0.25em]
      \toprule
	  \multicolumn{2}{c}{\textbf{Queues}} \\
      \bottomrule 
       \addlinespace[0.25em]
	  Vertex (2x)	 &	16 entries, 136 bytes/entry \\
	  Triangle, Tile &	16 entries, 388 bytes/entry \\
	  Fragment	 &	64 entries, 233 bytes/entry \\
	   \addlinespace[0.25em]
      \toprule
	  \multicolumn{2}{c}{\textbf{Caches}} \\
      \bottomrule
       \addlinespace[0.25em]
	  Vertex Cache	&	64 bytes/line, 2-way associative, \newline 4 KB, 1 bank, 1 cycle \\
	  Texture Caches (4x)	&	64 bytes/line, 2-way associative, \newline8 KB, 1 bank, 1 cycle \\
	  Tile Cache	&	64 bytes/line, 8-way associative, \newline 128 KB, 8 banks, 1 cycle \\
	  L2 Cache	& 64 bytes/line, 8-way associative, \newline 256 KB, 8 banks, 2 cycles \\
	  Color Buffer & 64 bytes/line, 1-way associative, \newline 1 KB, 1 bank, 1 cycle \\
	  Depth Buffer & 64 bytes/line, 1-way associative, \newline 1 KB, 1 bank, 1 cycle \\
	   \addlinespace[0.25em]
      \toprule
	  \multicolumn{2}{c}{\textbf{Non-programmable stages}} \\
      \bottomrule  
       \addlinespace[0.25em]
	  Primitive assembly	&	1 triangle/cycle \\
	  Rasterizer 	&	16 attributes/cycle \\
	  Early Z test	&	32 in-flight quad-fragments\\
	   \addlinespace[0.25em]
      \toprule
	  \multicolumn{2}{c}{\textbf{Programmable stages}} \\
      \bottomrule
       \addlinespace[0.25em]
	  Vertex Processor	&	1 vertex processor \\
	  Fragment Processor	&	4 fragment processors \\
	   \addlinespace[0.25em]
	  \toprule
	  \multicolumn{2}{c}{\textbf{Additional hardware}} \\
	   \bottomrule
	    \addlinespace[0.25em]
	  Sample Rate Table & 8100 entries, 4 bits/entry \\
      \bottomrule
	\end{tabular}
	\end{adjustbox}
\end{table}

\section{Results}\label{sec:results}
\begin{figure}[!htb]
    \centering
\includegraphics[width=.47\textwidth]{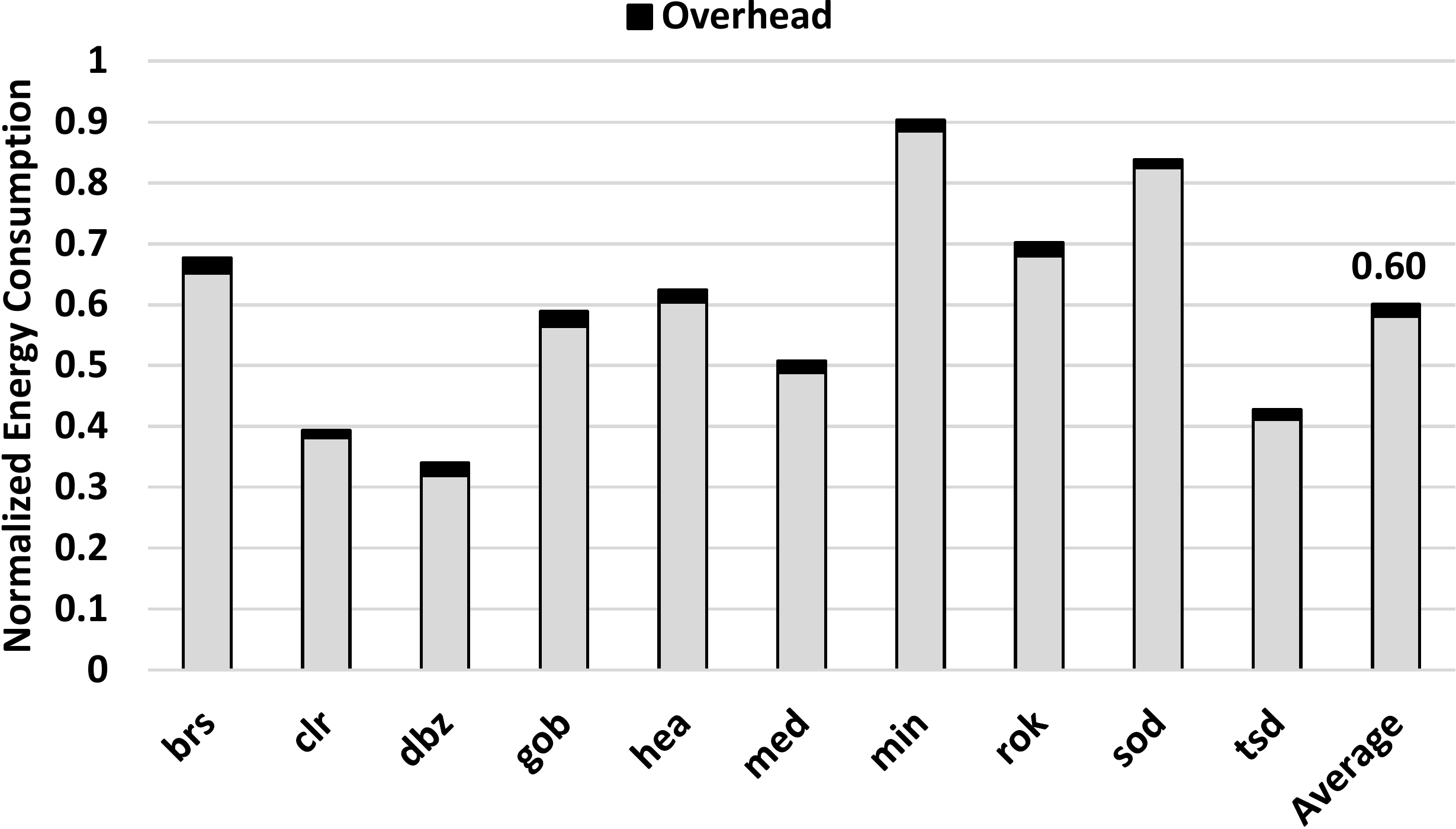}
    \caption{Energy consumption of DSR compared to the Baseline GPU.}
    \label{fig:energy}
\end{figure}

In this section we present the energy and performance gains of our proposal compared to those of the baseline GPU.
Figure \ref{fig:energy} shows the energy consumption of the whole system (GPU plus memory) with our DSR proposal normalized to the Baseline described in Section \ref{sec:background}. We can see that having independent and dynamic sampling rates for each tile achieves an average 40\% reduction of energy, with savings up to 67\% (for \textit{dbz}). Figure \ref{fig:energy} also shows the minor costs of activating DSR: the static and dynamic energy consumption of the Sampling Rate Table, and the logic and temporary memory required to compute the 2D DCT of the tiles (Figure \ref{fig:fascheme}). All together, they represent less than 2\% of the total energy consumption and less than 1\% of the area of the baseline GPU.

\begin{figure*}[!htb]
    \centering
\includegraphics[height=65mm, width=0.7\textwidth]{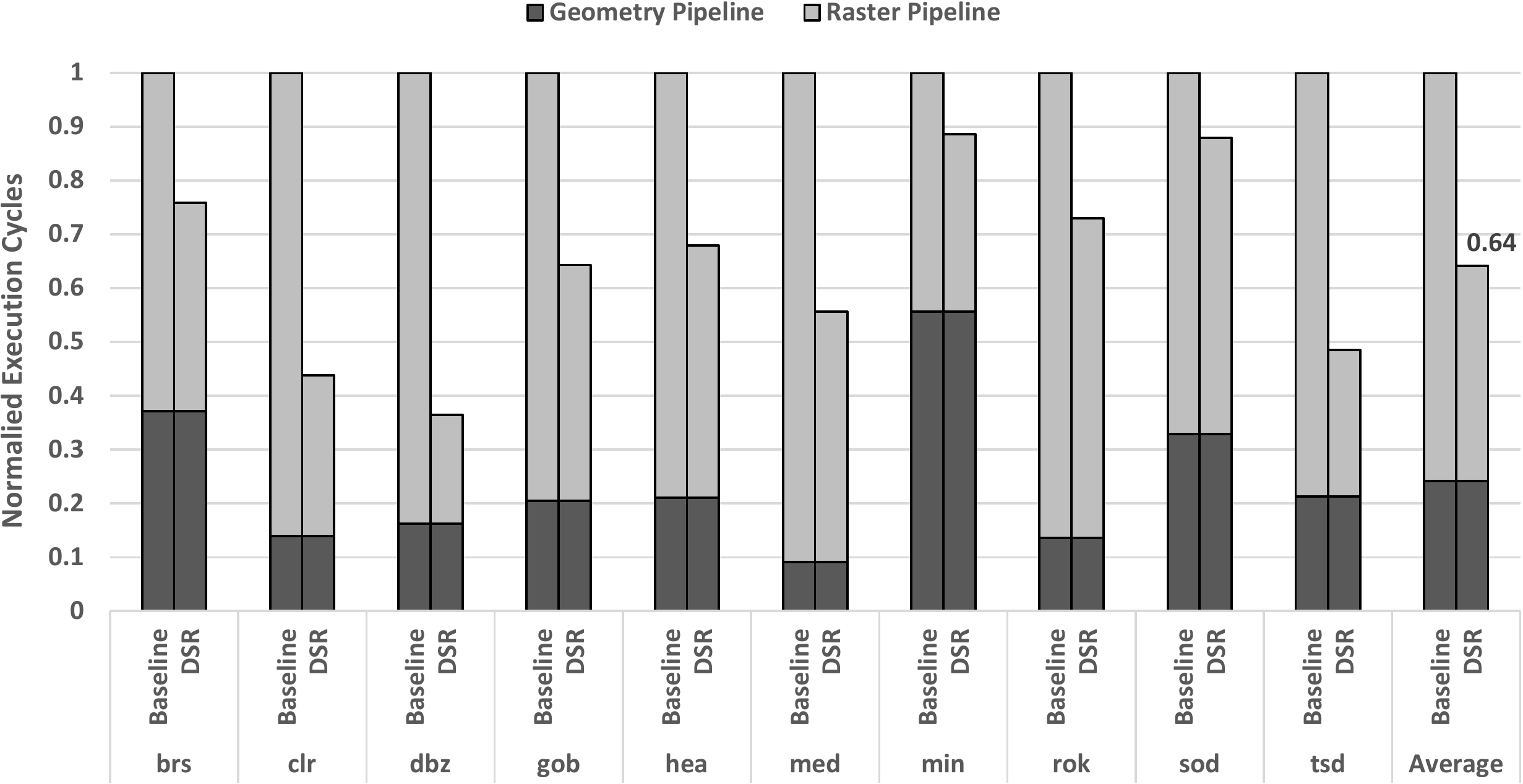}
    \caption{Execution time of DSR compared to the Baseline GPU.}
    \label{fig:speedup}
\end{figure*}

Figure \ref{fig:speedup} shows the reduction in execution cycles of DSR normalized to the Baseline design and broken down into Geometry and Raster cycles. On average, our proposal leads to 1.9x speedup in the Raster Pipeline, with maximums of more than 4x (\textit{dbz}). This translates into a 36\% global execution time reduction, since the Geometry Pipeline contains no modifications with respect to the baseline. Note that we do not incur in any execution time penalty, as the frequency analysis of the tiles and their sampling rate determination is completely overlapped with the Raster Pipeline activity.

\begin{figure}[!htb]
    \centering
\includegraphics[width=.47\textwidth]{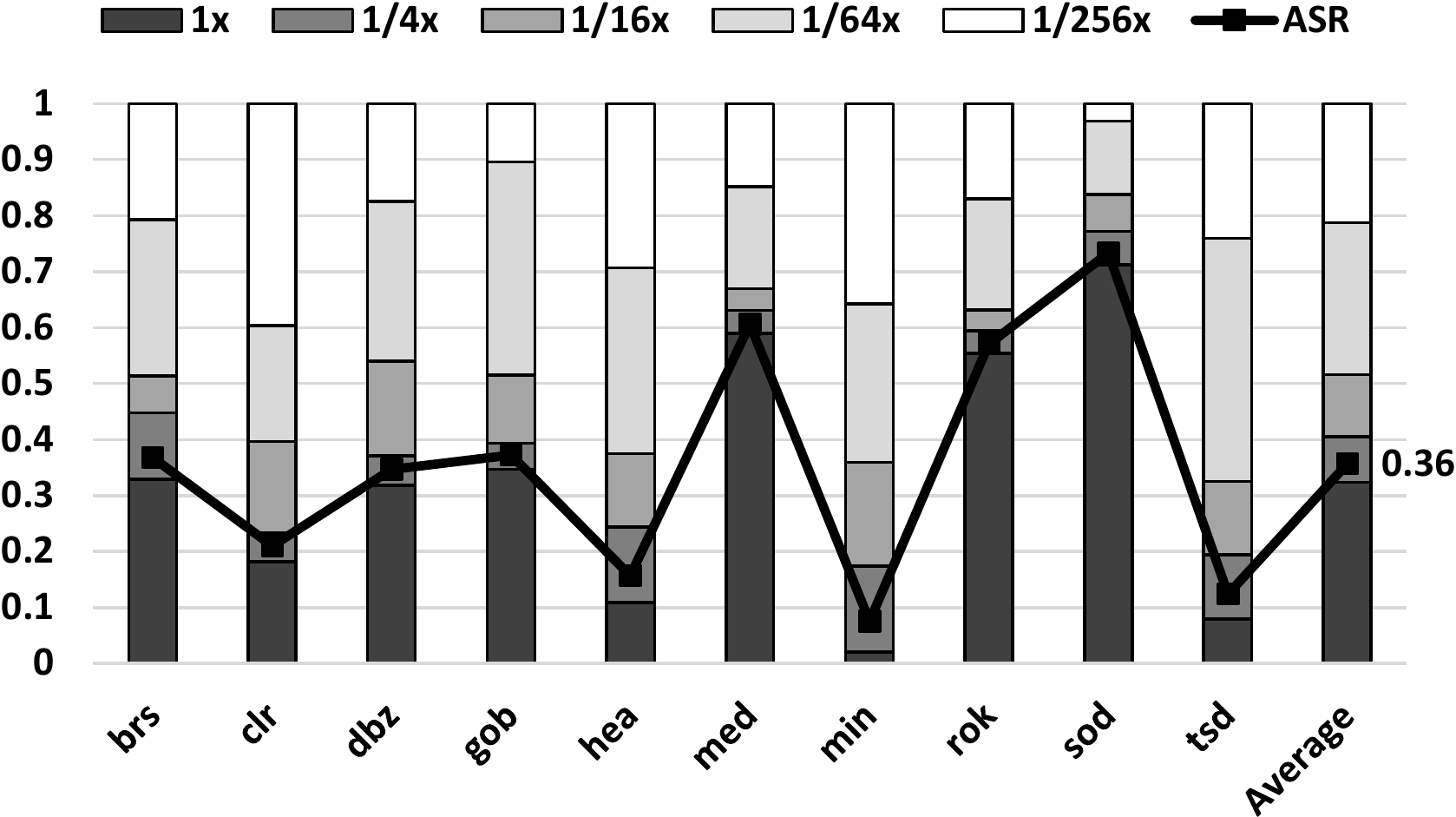}
    \caption{Breakdown of sampling rates.}
    \label{fig:breakdown}
\end{figure}

\begin{figure}[!htb]
    \centering
\includegraphics[width=.47\textwidth]{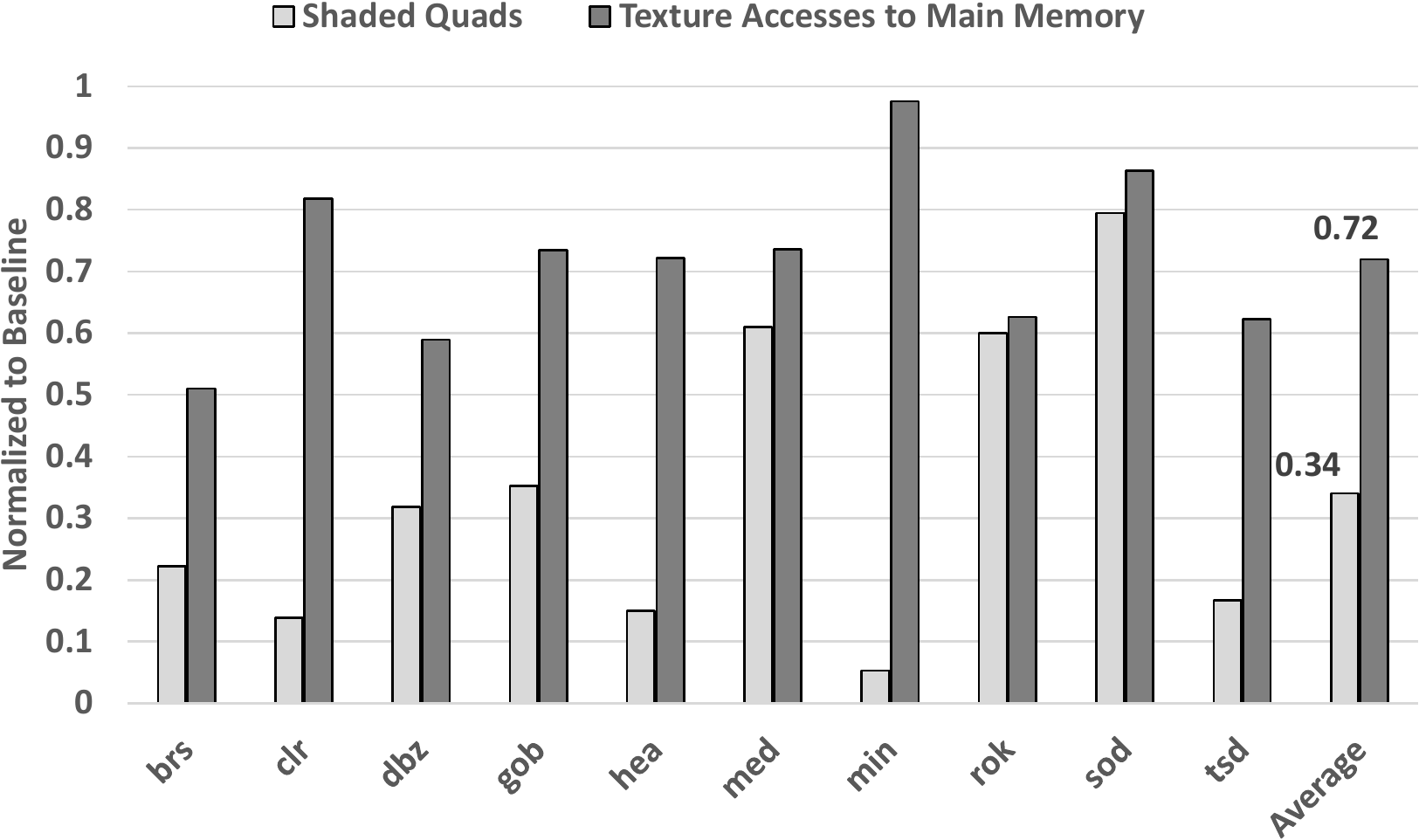}
    \caption{Shader activity of DSR compared to the Baseline GPU.}
    \label{fig:activity}
\end{figure}

The benefits in energy consumption and execution time of DSR are caused by sampling most tiles at lower rates, as shown in Figure \ref{fig:breakdown}. On average, less than half of the tiles are sampled at the baseline rate, while almost 40\% of the tiles are processed using the two lowest sampling rates (1/64x and 1/256x). The Average Sample Rate across all benchmarks and frames is thus reduced to 0.36 samples per fragment. This greatly reduces the activity of the Fragment Shaders, as shown in Figure \ref{fig:activity}. DSR reduces the average number of processed fragments by 66\% and the number of texture accesses to main memory by 28\% when compared to the Baseline. The gap between both numbers is caused by an increase in sparsity: as samples are taken at larger intervals, the likelihood of reusing a texture cache line is smaller than in the Baseline. However, the great reduction in processed fragments still allows for significant savings in overall texture traffic. 

\begin{figure}[!htb]
    \centering
\includegraphics[width=.47\textwidth]{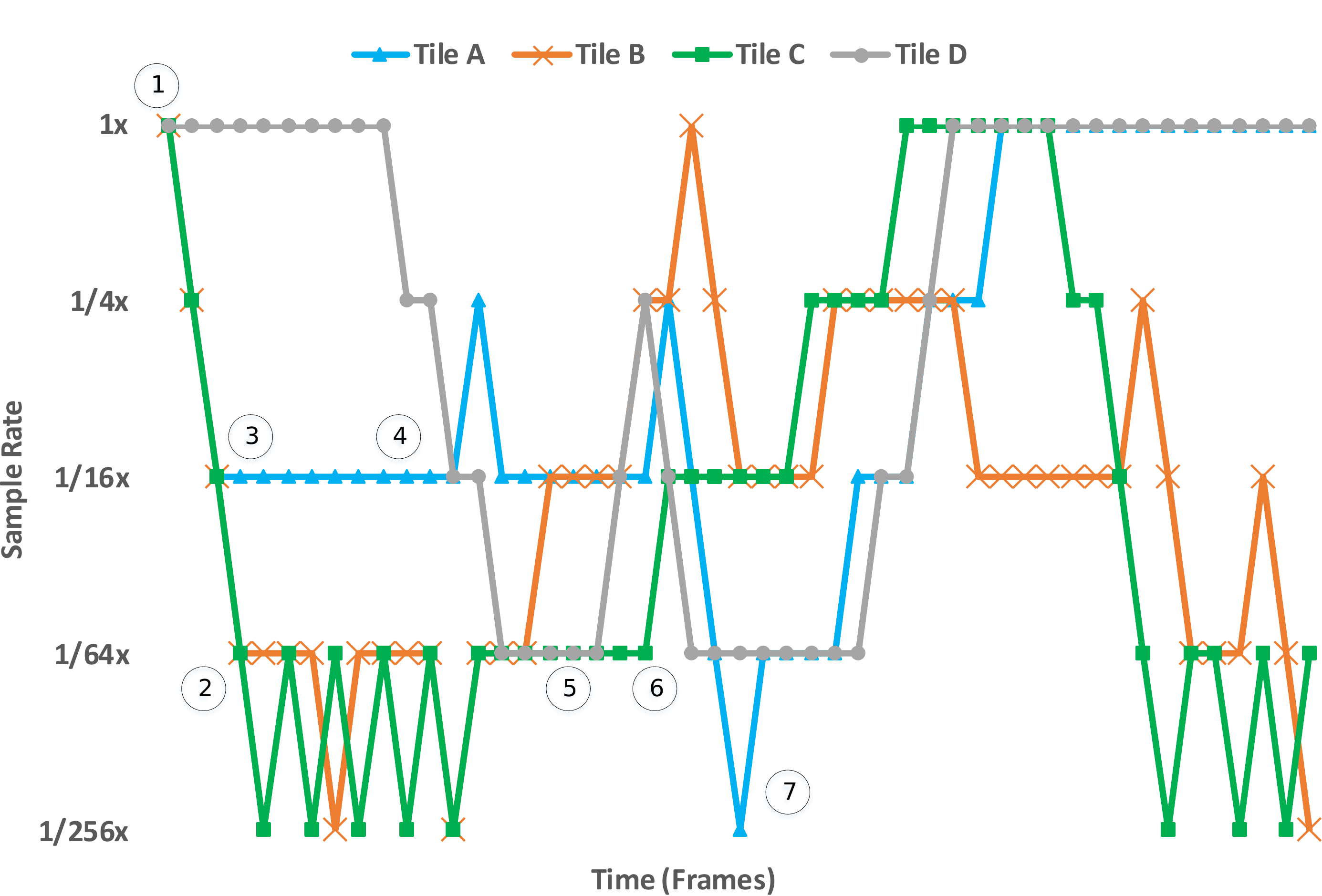}
    \caption{Evolution of the sampling rate of 4 different tiles over several frames.}
    \label{fig:srtiles}
\end{figure}

Rendered scenes in real-time applications tend to smoothly vary across consecutive frames. Therefore, the sampling rate requirements of tiles may evolve over time. As DSR analyzes the frequencies of the scene after rendering each tile, it manages to dynamically capture such changes and quickly adjust the sampling rates of all individual tiles accordingly. Figure \ref{fig:srtiles} illustrates this process by depicting the sampling rate of 4 tiles over several frames, starting at the beginning of the execution of the application. We can observe that DSR starts sampling all the tiles at the maximum rate, 1x (\circled{1}). Tiles A,B and C can be sampled at a much lower sampling rate, and spend a small transitory period of time continuously reducing their sampling rate (\circled{2}) until their optimal rate for their current spatial frequency is found (\circled{3}). Tiles remain in their estimated-optimal sampling rates (e.g., \circled{4}, \circled{5}) until the spatial frequencies in them change (e.g., \circled{6}). Note how every time that a tile is sampled at 1/256x rate (e.g., \circled{7}), the sampling rate is immediately increased in the following frame, as described in Figure \ref{fig:fsm}. It can be observed that the scene is not sampled uniformly neither in space (in a particular frame the sampling rate of the 4 tiles is normally different) nor in time (the sampling rate of a particular tile changes across the frames).

In our experiments, DSR has not produced a single error in all the generated frames, i.e., has not rendered any frame with a MSSIM lower than 95 when compared with the frame rendered at baseline sampling rate. Despite our benchmarks containing swift camera movements and object displacements across the screen, the similarity between consecutive frames allows the reuse of the estimated-best sampling rates without producing any visual artifacts. Albeit a sparse phenomenon, more abrupt alterations may occur in a particular frame, such as in a change of scene. The correctness of DSR cannot be guaranteed in these rare scenarios, as it is based on frame coherency. We have performed an experiment to quantify the effect that DSR has on image quality whenever there is a scene change. To do so, three additional 100-frame traces for the benchmarks listed in Table \ref{tab:benchmarks_changescene} have been generated. Each trace contains two different scene changes, emulated by entering and exiting the pause or settings menu of the application. With the renderization of these applications' frames with DSR active, we have observed that only the frame rendered immediately after each of the six scene changes is erroneous. Subsequent frames are indistinguishable from frames rendered at baseline sampling rate. It is well documented that the human eye requires some time to construe visual information: at least 40ms are required to perceive motion \cite{herzog2016time} and scenes cannot be properly recreated in less than 67ms \cite{goldstein2016sensation}. These times are greater than what a single frame lasts in 30 frames per second, the frame rate which is considered to be the minimum acceptable \cite{janzen201460,debattista2018frame}, so we can conclude that these potential errors affect only a single frame and will not be perceived by the user.

\begin{table}
\centering
\caption{Additional benchmarks for the image quality experiment.}
\label{tab:benchmarks_changescene}
\begin{tabular}{@{}lll@{}}
\toprule
\textbf{Benchmark}      & \textbf{Genre}  \\
\midrule
Alto's Odyssey   & Endless Runner\\
PlayerUnknown's Battlegrounds   & Battle Royale\\
Homescapes  & Puzzle\\
\bottomrule
\end{tabular}
\end{table}

\section{Related Work} \label{sec:rw}
There is a lot of interest in the graphics community in reducing shading costs so that more complex and realistic scenes can be rendered. Several techniques dynamically detect regions of the screen that can be sampled at lower rates by adding additional pipeline stages before or after the shading process. Deferred Adaptive Compute Shading \cite{mallett2018deferred} divides the framebuffer into levels, subsets of pixels progressively farther apart. Fragments are only shaded if the neighbor pixels from the previous level are not similar. Otherwise, the color of the fragment is computed by averaging the results of its neighbors. The work of Sathe and Akenine-M\"{o}ller \cite{sathe2015pixel} reuses shading computations for triangles that share an edge by adding a comparison queue before the processing stage. In Adaptive Image-Space Sampling \cite{stengel2016adaptive}, the resolution is reduced in areas that contain less perceivable detail, which are evaluated with an additional pass after the geometry processing. Conversely, DSR is architected to not introduce any time overhead by completely overlapping the sampling rate estimation for a tile with the rendering of the next one.

To avoid the runtime overhead of determining components with less detail, several works allow the programmer to statically determine the sampling rate. In coarse Pixel Shading  \cite{vaidyanathan2014coarse}, the sample rate of each primitive can be controlled based on their vertex attributes. He et al. \cite{he2014extending} design new language abstractions that grant each shader program  the ability to determine which components can be processed at which rate. In NVIDIA's Variable Rate Shading \cite{nvidiavrs}, the programmer decides which sampling rate to apply in each section of the screen. DSR, on the other hand dynamically estimates the best sampling rates in each tile by using a hardware-only mechanism, in a completely transparent manner to the programmer.

Frame coherence has been previously leveraged to reduce the number of samples to process. In Checkerboard rendering \cite{mcferroncheckerboard}, each frame shades an alternate half of the pixels in the screen. The color of the non-shaded half is obtained by applying reconstruction filters to the results obtained in the preceding frame. 
A large number of shading computations are avoided at the cost of some visual artifacts, since the lossy nature of the reconstruction and the fixed undersampling cannot perfectly reproduce neither motion nor visibility changes. In contrast, DSR estimates sampling rates at the finer granularity of tiles, can render tiles at the small rate of only one fragment per tile and does not affect image quality because it only reduces the sampling rate whenever a tile does not contain high spatial frequencies.

\section{Conclusions} \label{sec:conclusions}
This paper proposes Dynamic Sampling Rate (DSR), a novel microarchitectural technique to reduce shader executions by determining the lowest sampling rate for each tile in a frame that does not reduce the overall quality of the rendered images. DSR analyzes the frequency components of a tile once it has been processed and decides the rate in which the tile's triangles will be sampled in the following frame. The sampling rate prediction leverages the frame-to-frame coherence inherent in animated graphics applications, which results in a high likelihood that the frequency components of a tile are maintained across consecutive frames.

We have shown that for a set of unmodified commercial Android applications DSR reduces the fragment-level redundancy by 66\% on average with minimal hardware overhead, leading to an average speedup of 1.68x and energy savings of 40\%.


\ifCLASSOPTIONcompsoc
  \section*{Acknowledgments}
\else
  \section*{Acknowledgment}
\fi

This work has been supported by the the CoCoUnit ERC Advanced Grant of the EU’s Horizon 2020 program (grant No 833057), Spanish State Research Agency (MCIN/AEI) under grant PID2020-113172RB-I00, the ICREA Academia program, and the Generalitat de Catalunya under grant FI-DGR 2016

\ifCLASSOPTIONcaptionsoff
  \newpage
\fi




\bibitem{ahmed1974discrete}
N.~Ahmed, T.~Natarajan, and K.~R. Rao, ``Discrete cosine transform,''
  \emph{IEEE transactions on Computers}, vol. 100, no.~1, pp. 90--93, 1974.

\bibitem{akeley1993reality}
K.~Akeley, ``Reality engine graphics,'' in \emph{Proceedings of the 20th annual
  conference on Computer graphics and interactive techniques}.\hskip 1em plus
  0.5em minus 0.4em\relax ACM, 1993, pp. 109--116.

\bibitem{akenine2008graphics}
T.~Akenine-Moller and J.~Strom, ``Graphics processing units for handhelds,''
  \emph{Proceedings of the IEEE}, vol.~96, no.~5, pp. 779--789, 2008.

\bibitem{anandtechGPU}
(accessed October, 2020) Qualcomm snapdragon s4 (krait) performance preview.
  \url{http://www.anandtech.com/show/5559/qualcomm-snapdragon-s4-krait-performance-preview-msm8960-adreno-225-benchmarks/4}.

\bibitem{8675248}
M.~{Anglada}, E.~{de Lucas}, J.~{Parcerisa}, J.~L. {Aragón}, P.~{Marcuello},
  and A.~{González}, ``Rendering elimination: Early discard of redundant tiles
  in the graphics pipeline,'' in \emph{2019 IEEE International Symposium on
  High Performance Computer Architecture (HPCA)}, Feb 2019, pp. 623--634.

\bibitem{mali450}
(accessed October 2020) Mali-450 gpu.
  \url{https://developer.arm.com/products/graphics-and-multimedia/mali-gpus/mali-450-gpu}.

\bibitem{arnau2013teapot}
J.-M. Arnau, J.-M. Parcerisa, and P.~Xekalakis, ``Teapot: a toolset for
  evaluating performance, power and image quality on mobile graphics systems,''
  in \emph{Proceedings of the 27th International ACM Conference on
  Supercomputing}.\hskip 1em plus 0.5em minus 0.4em\relax ACM, 2013, pp.
  37--46.

\bibitem{enriqueTesis}
E.~de~Lucas, ``Reducing redundancy of real time computer graphics in mobile
  systems,'' Ph.D. dissertation, UPC, Computer Architecture Department, 2018.

\bibitem{debattista2018frame}
K.~Debattista, K.~Bugeja, S.~Spina, T.~Bashford-Rogers, and V.~Hulusic, ``Frame
  rate vs resolution: A subjective evaluation of spatiotemporal perceived
  quality under varying computational budgets,'' in \emph{Computer Graphics
  Forum}, vol.~37, no.~1.\hskip 1em plus 0.5em minus 0.4em\relax Wiley Online
  Library, 2018, pp. 363--374.

\bibitem{flynn2013image}
J.R.~Flynn, S.~Ward, M.~Eisemann, J.~Abich, D.~Poole, ``Image quality assessment using the ssim and the just noticeable difference paradigm,'' in \emph{International Conference on Engineering Psychology and Cognitive Ergonomics}\hskip 1em plus 0.5em minus 0.4em\relax Springer, 2013, pp. 23--30.



\bibitem{gao2009image}
X.~Gao, W.~Lu, T.~Dacheng, X.~Li, ``Image quality assessment based on multiscale geometric analysis,''
  \emph{IEEE Transactions on Image Processing}, vol.~18, no.~7, pp. 1409--1423, 2009. 

\bibitem{gapid}
(accessed October 2020) Gapid.
  \url{https://developers.google.com/vr/develop/unity/gapid}.

\bibitem{goldstein2016sensation}
E.~B. Goldstein and J.~Brockmole, \emph{Sensation and perception}.\hskip 1em
  plus 0.5em minus 0.4em\relax Cengage Learning, 2016.

\bibitem{googleplay}
(accessed October 2020) Google play. \url{https://play.google.com}.

\bibitem{he2014extending}
Y.~He, Y.~Gu, and K.~Fatahalian, ``Extending the graphics pipeline with
  adaptive, multi-rate shading,'' \emph{ACM Transactions on Graphics (TOG)},
  vol.~33, no.~4, p. 142, 2014.

\bibitem{herzog2016time}
M.~H. Herzog, T.~Kammer, and F.~Scharnowski, ``Time slices: what is the
  duration of a percept?'' \emph{PLoS biology}, vol.~14, no.~4, p. e1002433,
  2016.

\bibitem{Hubschman1982}
H.~Hubschman \emph{et~al.}, ``Frame-to-frame coherence and the hidden surface
  computation: constraints for a convex world,'' \emph{ACM Trans. on Graphics},
  vol.~1, no.~2, pp. 129--162, 1982.

\bibitem{janzen201460}
B.~F. Janzen and R.~J. Teather, ``Is 60 fps better than 30?: the impact of
  frame rate and latency on moving target selection,'' in \emph{Proceedings of
  the extended abstracts of the 32nd annual ACM conference on Human factors in
  computing systems}.\hskip 1em plus 0.5em minus 0.4em\relax ACM, 2014, pp.
  1477--1482.


\bibitem{ma2010new}
Q.~Ma, L.~Zhang, B.~Wang, ``New strategy for image and video quality assessment,''
  \emph{Journal of Electronic Imaging}, vol.~19, no.~1, 2010.


\bibitem{mallett2018deferred}
I.~Mallett and C.~Yuksel, ``Deferred adaptive compute shading,'' in
  \emph{Proceedings of the Conference on High-Performance Graphics}.\hskip 1em
  plus 0.5em minus 0.4em\relax ACM, 2018, p.~3.

\bibitem{maule2012transparency}
M.~Maule, J.~L. Comba, R.~Torchelsen, and R.~Bastos, ``Transparency and
  anti-aliasing techniques for real-time rendering,'' in \emph{2012 25th
  SIBGRAPI Conference on Graphics, Patterns and Images Tutorials}.\hskip 1em
  plus 0.5em minus 0.4em\relax IEEE, 2012, pp. 50--59.

\bibitem{mcferroncheckerboard}
(accessed October 2020) Checkerboard rendering for real-time upscaling on intel{\textregistered} integrated graphics. 
  \url{https://software.intel.com/en-us/articles/checkerboard-rendering-for-real-time-upscaling-on-intel-integrated-graphics}

\bibitem{nvidiavrs}
(accessed October 2020) Turing gpu architecture.
  \url{https://www.nvidia.com/content/dam/en-zz/Solutions/design-visualization/technologies/turing-architecture/NVIDIA-Turing-Architecture-Whitepaper.pdf}.

\bibitem{nyquist1928certain}
H.~Nyquist, ``Certain topics in telegraph transmission theory,''
  \emph{Transactions of the American Institute of Electrical Engineers},
  vol.~47, no.~2, pp. 617--644, 1928.

\bibitem{patil2015characterization}
S.~Patil, Y.~Kim, K.~Korgaonkar, I.~Awwal, and T.~S. Rosing, ``Characterization
  of user’s behavior variations for design of replayable mobile workloads,''
  in \emph{International Conference on Mobile Computing, Applications, and
  Services}.\hskip 1em plus 0.5em minus 0.4em\relax Springer, 2015, pp. 51--70.

\bibitem{pool2012energy}
J.~Pool, ``Energy-precision tradeoffs in the graphics pipeline,'' Ph.D.
  dissertation, The University of North Carolina at Chapel Hill, 2012.

\bibitem{rao2014discrete}
K.~R. Rao and P.~Yip, \emph{Discrete cosine transform: algorithms, advantages,
  applications}.\hskip 1em plus 0.5em minus 0.4em\relax Academic press, 2014.

\bibitem{sathe2015pixel}
R.~Sathe and T.~Akenine-M{\"o}ller, ``Pixel merge unit,'' in \emph{Eurographics
  (Short Papers)}, 2015, pp. 53--56.

\bibitem{shebanow2013evolution}
M.~Shebanow, ``An evolution of mobile graphics,'' \emph{Keynote talk at High
  Performance Graphics}, 2013.

\bibitem{sihvo2005row}
T.~Sihvo and J.~Niittylahti, ``Row-column decomposition based 2d transform
  optimization on subword parallel processors,'' in \emph{International
  Symposium on Signals, Circuits and Systems, 2005. ISSCS 2005.}, vol.~1.\hskip
  1em plus 0.5em minus 0.4em\relax IEEE, 2005, pp. 99--102.

\bibitem{stengel2016adaptive}
M.~Stengel, S.~Grogorick, M.~Eisemann, and M.~Magnor, ``Adaptive image-space
  sampling for gaze-contingent real-time rendering,'' in \emph{Computer
  Graphics Forum}, vol.~35, no.~4.\hskip 1em plus 0.5em minus 0.4em\relax Wiley
  Online Library, 2016, pp. 129--139.

\bibitem{dwdct}
(accessed October 2020) Designware 2d dct.
  \url{https://www.synopsys.com/dw/ipdir.php?c=DW_dct_2d}.

\bibitem{synopsys}
(accessed October 2020) Synopsys. \url{https://www.synopsys.com}.

\bibitem{vaidyanathan2014coarse}
K.~Vaidyanathan, M.~Salvi, R.~Toth, T.~Foley, T.~Akenine-M{\"o}ller,
  J.~Nilsson, J.~Munkberg, J.~Hasselgren, M.~Sugihara, P.~Clarberg
  \emph{et~al.}, ``Coarse pixel shading,'' in \emph{Proceedings of High
  Performance Graphics}.\hskip 1em plus 0.5em minus 0.4em\relax Eurographics
  Association, 2014, pp. 9--18.

\bibitem{gallium}
(accessed October 2020) Gallium3d. \url{https://www.freedesktop.org/wiki/Software/gallium}.

\bibitem{wang2004image}
Z.~Wang, A.~C. Bovik, H.~R. Sheikh, E.~P. Simoncelli \emph{et~al.}, ``Image
  quality assessment: from error visibility to structural similarity,''
  \emph{IEEE transactions on image processing}, vol.~13, no.~4, pp. 600--612,
  2004.


\begin{thebibliography}{36}


\bibitem{ahmed1974discrete}
N.~Ahmed, T.~Natarajan, and K.~R. Rao, ``Discrete cosine transform,''
  \emph{IEEE transactions on Computers}, vol. 100, no.~1, pp. 90--93, 1974.

\bibitem{akeley1993reality}
K.~Akeley, ``Reality engine graphics,'' in \emph{Proceedings of the 20th annual
  conference on Computer graphics and interactive techniques}.\hskip 1em plus
  0.5em minus 0.4em\relax ACM, 1993, pp. 109--116.

\bibitem{akenine2008graphics}
T.~Akenine-Moller and J.~Strom, ``Graphics processing units for handhelds,''
  \emph{Proceedings of the IEEE}, vol.~96, no.~5, pp. 779--789, 2008.

\bibitem{anandtechGPU}
(accessed October, 2020) Qualcomm snapdragon s4 (krait) performance preview.
  \url{http://www.anandtech.com/show/5559/qualcomm-snapdragon-s4-krait-performance-preview-msm8960-adreno-225-benchmarks/4}.

\bibitem{8675248}
M.~{Anglada}, E.~{de Lucas}, J.~{Parcerisa}, J.~L. {Aragón}, P.~{Marcuello},
  and A.~{González}, ``Rendering elimination: Early discard of redundant tiles
  in the graphics pipeline,'' in \emph{2019 IEEE International Symposium on
  High Performance Computer Architecture (HPCA)}, Feb 2019, pp. 623--634.

\bibitem{mali450}
(accessed October 2020) Mali-450 gpu.
  \url{https://developer.arm.com/products/graphics-and-multimedia/mali-gpus/mali-450-gpu}.

\bibitem{arnau2013teapot}
J.-M. Arnau, J.-M. Parcerisa, and P.~Xekalakis, ``Teapot: a toolset for
  evaluating performance, power and image quality on mobile graphics systems,''
  in \emph{Proceedings of the 27th International ACM Conference on
  Supercomputing}.\hskip 1em plus 0.5em minus 0.4em\relax ACM, 2013, pp.
  37--46.

\bibitem{enriqueTesis}
E.~de~Lucas, ``Reducing redundancy of real time computer graphics in mobile
  systems,'' Ph.D. dissertation, UPC, Computer Architecture Department, 2018.

\bibitem{debattista2018frame}
K.~Debattista, K.~Bugeja, S.~Spina, T.~Bashford-Rogers, and V.~Hulusic, ``Frame
  rate vs resolution: A subjective evaluation of spatiotemporal perceived
  quality under varying computational budgets,'' in \emph{Computer Graphics
  Forum}, vol.~37, no.~1.\hskip 1em plus 0.5em minus 0.4em\relax Wiley Online
  Library, 2018, pp. 363--374.

\bibitem{flynn2013image}
J.R.~Flynn, S.~Ward, M.~Eisemann, J.~Abich, D.~Poole, ``Image quality assessment using the ssim and the just noticeable difference paradigm,'' in \emph{International Conference on Engineering Psychology and Cognitive Ergonomics}\hskip 1em plus 0.5em minus 0.4em\relax Springer, 2013, pp. 23--30.



\bibitem{gao2009image}
X.~Gao, W.~Lu, T.~Dacheng, X.~Li, ``Image quality assessment based on multiscale geometric analysis,''
  \emph{IEEE Transactions on Image Processing}, vol.~18, no.~7, pp. 1409--1423, 2009. 

\bibitem{gapid}
(accessed October 2020) Gapid.
  \url{https://developers.google.com/vr/develop/unity/gapid}.

\bibitem{goldstein2016sensation}
E.~B. Goldstein and J.~Brockmole, \emph{Sensation and perception}.\hskip 1em
  plus 0.5em minus 0.4em\relax Cengage Learning, 2016.

\bibitem{googleplay}
(accessed October 2020) Google play. \url{https://play.google.com}.

\bibitem{he2014extending}
Y.~He, Y.~Gu, and K.~Fatahalian, ``Extending the graphics pipeline with
  adaptive, multi-rate shading,'' \emph{ACM Transactions on Graphics (TOG)},
  vol.~33, no.~4, p. 142, 2014.

\bibitem{herzog2016time}
M.~H. Herzog, T.~Kammer, and F.~Scharnowski, ``Time slices: what is the
  duration of a percept?'' \emph{PLoS biology}, vol.~14, no.~4, p. e1002433,
  2016.

\bibitem{Hubschman1982}
H.~Hubschman \emph{et~al.}, ``Frame-to-frame coherence and the hidden surface
  computation: constraints for a convex world,'' \emph{ACM Trans. on Graphics},
  vol.~1, no.~2, pp. 129--162, 1982.

\bibitem{janzen201460}
B.~F. Janzen and R.~J. Teather, ``Is 60 fps better than 30?: the impact of
  frame rate and latency on moving target selection,'' in \emph{Proceedings of
  the extended abstracts of the 32nd annual ACM conference on Human factors in
  computing systems}.\hskip 1em plus 0.5em minus 0.4em\relax ACM, 2014, pp.
  1477--1482.


\bibitem{ma2010new}
Q.~Ma, L.~Zhang, B.~Wang, ``New strategy for image and video quality assessment,''
  \emph{Journal of Electronic Imaging}, vol.~19, no.~1, 2010.


\bibitem{mallett2018deferred}
I.~Mallett and C.~Yuksel, ``Deferred adaptive compute shading,'' in
  \emph{Proceedings of the Conference on High-Performance Graphics}.\hskip 1em
  plus 0.5em minus 0.4em\relax ACM, 2018, p.~3.

\bibitem{maule2012transparency}
M.~Maule, J.~L. Comba, R.~Torchelsen, and R.~Bastos, ``Transparency and
  anti-aliasing techniques for real-time rendering,'' in \emph{2012 25th
  SIBGRAPI Conference on Graphics, Patterns and Images Tutorials}.\hskip 1em
  plus 0.5em minus 0.4em\relax IEEE, 2012, pp. 50--59.

\bibitem{mcferroncheckerboard}
(accessed October 2020) Checkerboard rendering for real-time upscaling on intel{\textregistered} integrated graphics. 
  \url{https://software.intel.com/en-us/articles/checkerboard-rendering-for-real-time-upscaling-on-intel-integrated-graphics}

\bibitem{nvidiavrs}
(accessed October 2020) Turing gpu architecture.
  \url{https://www.nvidia.com/content/dam/en-zz/Solutions/design-visualization/technologies/turing-architecture/NVIDIA-Turing-Architecture-Whitepaper.pdf}.

\bibitem{nyquist1928certain}
H.~Nyquist, ``Certain topics in telegraph transmission theory,''
  \emph{Transactions of the American Institute of Electrical Engineers},
  vol.~47, no.~2, pp. 617--644, 1928.

\bibitem{patil2015characterization}
S.~Patil, Y.~Kim, K.~Korgaonkar, I.~Awwal, and T.~S. Rosing, ``Characterization
  of user’s behavior variations for design of replayable mobile workloads,''
  in \emph{International Conference on Mobile Computing, Applications, and
  Services}.\hskip 1em plus 0.5em minus 0.4em\relax Springer, 2015, pp. 51--70.

\bibitem{pool2012energy}
J.~Pool, ``Energy-precision tradeoffs in the graphics pipeline,'' Ph.D.
  dissertation, The University of North Carolina at Chapel Hill, 2012.

\bibitem{rao2014discrete}
K.~R. Rao and P.~Yip, \emph{Discrete cosine transform: algorithms, advantages,
  applications}.\hskip 1em plus 0.5em minus 0.4em\relax Academic press, 2014.

\bibitem{sathe2015pixel}
R.~Sathe and T.~Akenine-M{\"o}ller, ``Pixel merge unit,'' in \emph{Eurographics
  (Short Papers)}, 2015, pp. 53--56.

\bibitem{shebanow2013evolution}
M.~Shebanow, ``An evolution of mobile graphics,'' \emph{Keynote talk at High
  Performance Graphics}, 2013.

\bibitem{sihvo2005row}
T.~Sihvo and J.~Niittylahti, ``Row-column decomposition based 2d transform
  optimization on subword parallel processors,'' in \emph{International
  Symposium on Signals, Circuits and Systems, 2005. ISSCS 2005.}, vol.~1.\hskip
  1em plus 0.5em minus 0.4em\relax IEEE, 2005, pp. 99--102.

\bibitem{stengel2016adaptive}
M.~Stengel, S.~Grogorick, M.~Eisemann, and M.~Magnor, ``Adaptive image-space
  sampling for gaze-contingent real-time rendering,'' in \emph{Computer
  Graphics Forum}, vol.~35, no.~4.\hskip 1em plus 0.5em minus 0.4em\relax Wiley
  Online Library, 2016, pp. 129--139.

\bibitem{dwdct}
(accessed October 2020) Designware 2d dct.
  \url{https://www.synopsys.com/dw/ipdir.php?c=DW_dct_2d}.

\bibitem{synopsys}
(accessed October 2020) Synopsys. \url{https://www.synopsys.com}.

\bibitem{vaidyanathan2014coarse}
K.~Vaidyanathan, M.~Salvi, R.~Toth, T.~Foley, T.~Akenine-M{\"o}ller,
  J.~Nilsson, J.~Munkberg, J.~Hasselgren, M.~Sugihara, P.~Clarberg
  \emph{et~al.}, ``Coarse pixel shading,'' in \emph{Proceedings of High
  Performance Graphics}.\hskip 1em plus 0.5em minus 0.4em\relax Eurographics
  Association, 2014, pp. 9--18.

\bibitem{gallium}
(accessed October 2020) Gallium3d. \url{https://www.freedesktop.org/wiki/Software/gallium}.

\bibitem{wang2004image}
Z.~Wang, A.~C. Bovik, H.~R. Sheikh, E.~P. Simoncelli \emph{et~al.}, ``Image
  quality assessment: from error visibility to structural similarity,''
  \emph{IEEE transactions on image processing}, vol.~13, no.~4, pp. 600--612,
  2004.

  
  
 

 
  



\end{thebibliography}
%

%

\begin{IEEEbiography}[{\includegraphics[width=1in,height=1.25in,clip,keepaspectratio]{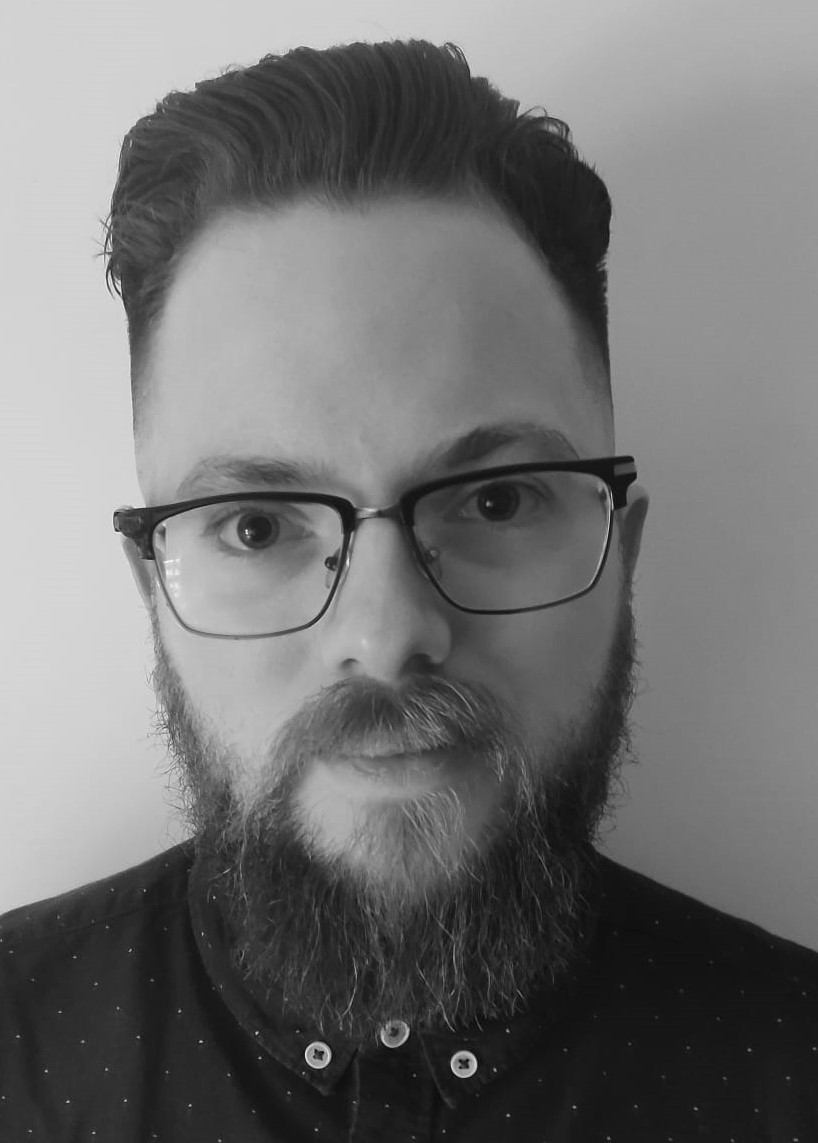}}]{Martí Anglada}
received his M.S. degree in High Performance Computing in 2015 and his PhD degree in Computer Architecture in 2020, both from Universitat Politècnica de Catalunya (UPC-BarcelonaTech). He joined the UPC-BarcelonaTech Architectures and Compilers research group in July 2014. His research is focused on energy-efficient architectures for mobile GPUs.
\end{IEEEbiography}

\begin{IEEEbiography}
[{\includegraphics[trim={18cm 20cm 18cm 20cm}, width=1in,height=1.25in,clip,keepaspectratio]{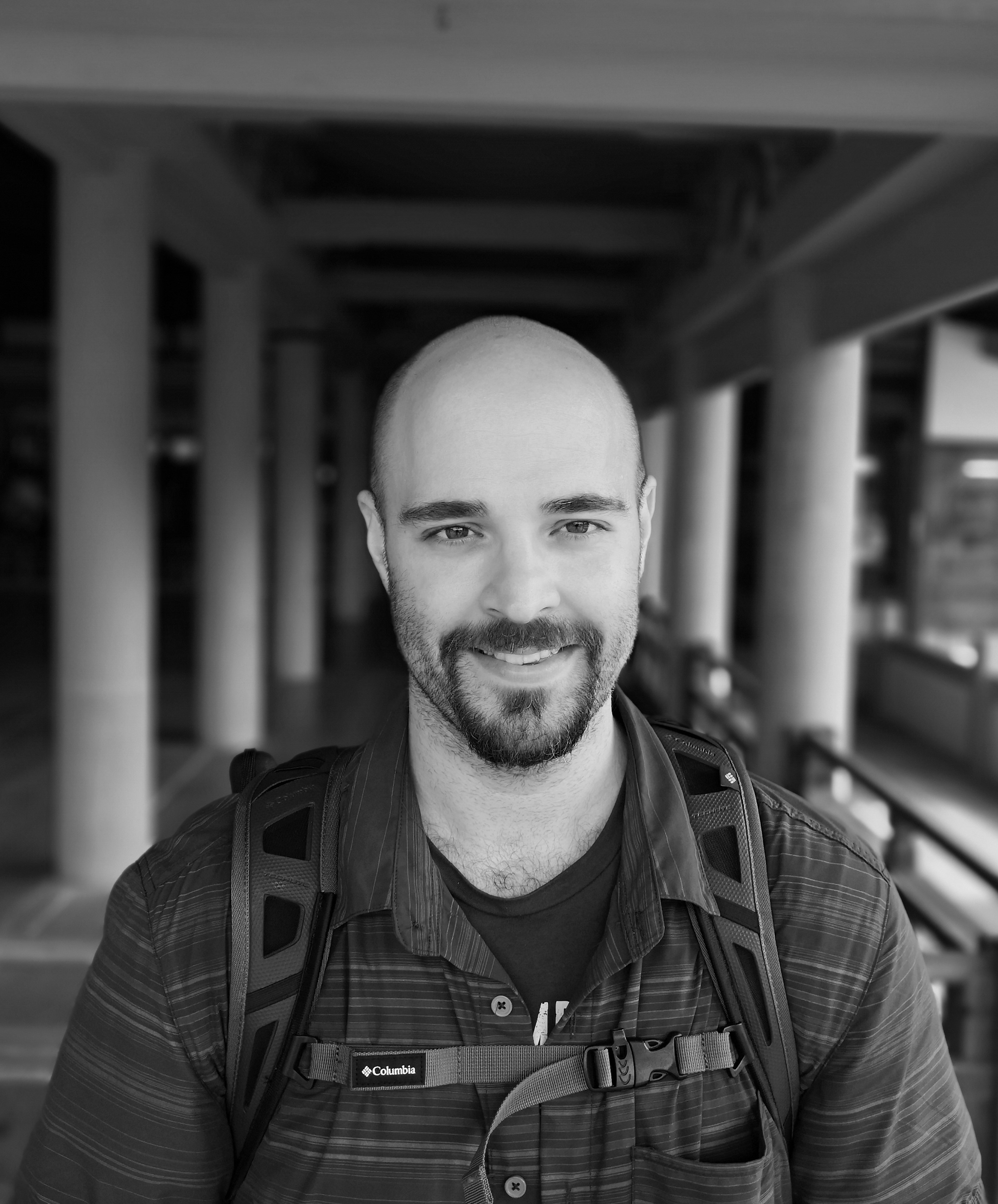}}] {Enrique de Lucas} received his B.S. degree in Computer Science in 2010 and M.S. degree in Computer Engineering in 2011 both from Complutense University of Madrid (UCM), Spain. During 2012 he worked on processor microarchitecture at Intel Labs. By February 2013 he joined ARCO research group of Universitat Politècnica de Catalunya (UPC), Barcelona (Spain), where he completed the PhD degree in Computer Architecture by 2018. His main research interests included techniques to exploit inter-frame coherency and reduce redundancy in the graphics subsystem for increasing the energy-efficiency of GPUs. By January 2017 he joined Esperanto Technologies, Barcelona (Spain), where he held a position in processor architecture and graphics with special focus in RISC-V development. Since March 2020 he is GPU Architect at Imagination Technologies, Kings Langley (United Kingdom).
\end{IEEEbiography}

\begin{IEEEbiography} [{\includegraphics[width=1in,height=1.25in,clip,keepaspectratio]{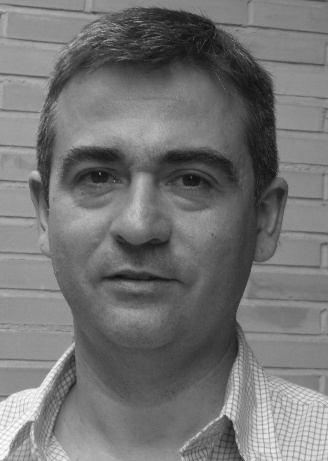}}]{Joan-Manuel Parcerisa}
Joan-Manuel Parcerisa received his M.S. and Ph.D. degrees in Computer Science from the Universitat Politècnica de Catalunya (UPC), in Barcelona, Spain, in 1993 and 2004 respectively.
Since 1994 he is a full time assistant professor at the Computer Architecture Department at the Universitat Politècnica de Catalunya. His research topics include ultra-low power GPU architectures for mobile devices, decoupled access/execute architectures, clustered microarchitectures, predication for OoO execution and cache memories.
\end{IEEEbiography}

\begin{IEEEbiography} [{\includegraphics[width=1in,height=1.25in,clip,keepaspectratio]{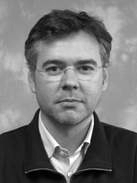}}] {Juan L. Aragón}
is an Associate Professor in Computer Architecture at the University of Murcia (UMU), Spain. In 2003 he received his PhD degree in Computer Engineering from the UMU, followed by a 1-year postdoc at the University of California, Irvine. He has also been a Visiting Researcher at EPFL (Lausanne, Switzerland) in 2013; and a Visiting Researcher at Princeton University (USA) in 2015, 2017, 2018 and 2019. Dr. Aragón has advised 4 PhD theses and has co-authored 50 research papers in major conferences and journals. His research interests are focused on computer architecture, with special emphasis on heterogeneous parallel systems, application-specific accelerators, microarchitecture design, and GPUs.
\end{IEEEbiography}

\begin{IEEEbiography} [{\includegraphics[width=1in,height=1.25in,clip,keepaspectratio]{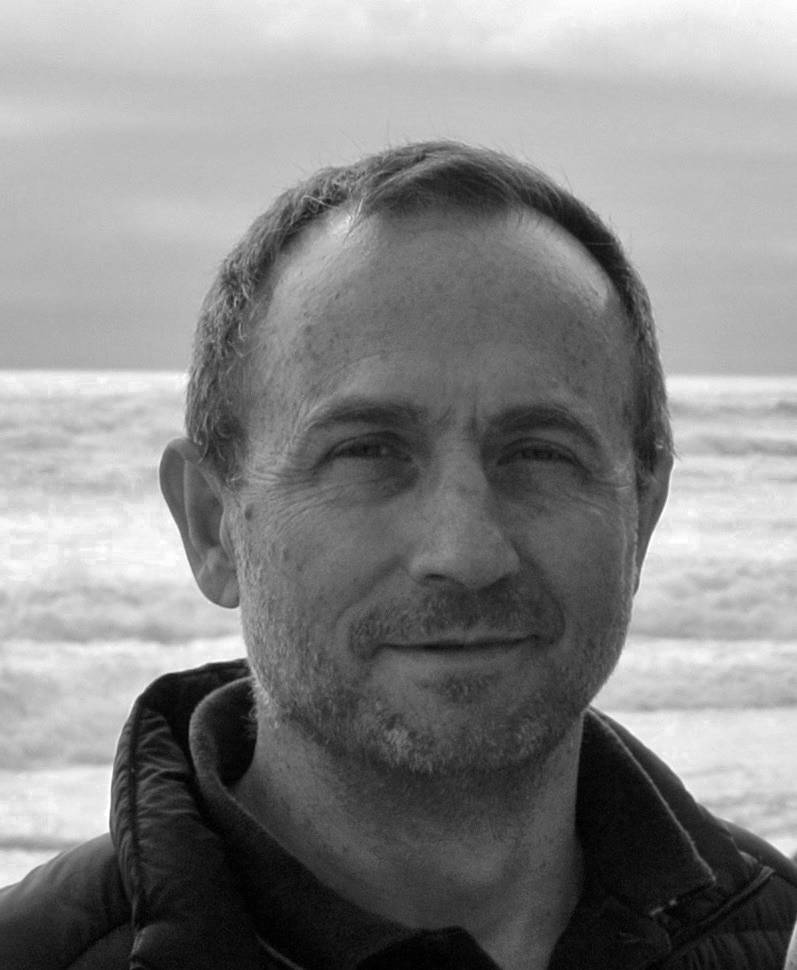}}] {Antonio González} (PhD 1989) is a Full Professor at the Computer Architecture Department of the Universitat Politècnica de Catalunya, Barcelona (Spain), and the director of the Architecture and Compilers research group. He was the founding director of the Intel Barcelona Research Center from 2002 to 2014. His research has focused on computer architecture and compilers, with a special emphasis on cognitive computing systems and graphics processors in recent years. He has published over 370 papers, and has served as associate editor of five IEEE and ACM journals, program chair for ISCA, MICRO, HPCA, ICS and ISPASS, general chair for MICRO and HPCA, and PC member for more than 130 symposia. He is an IEEE Fellow.

\end{IEEEbiography}





\end{document}